\documentclass[preprint]{aastex}
\usepackage{emulateapj5}
\usepackage{apjfonts}
\input psfig.tex
\slugcomment{ApJ Vol. 558, in press}
\shorttitle{Palomar Seyfert radio sources}
\shortauthors{Ulvestad \& Ho}
\def\H2{\ion{H}{2}}

\def\hii{\ion{H}{2}}
\def\oiii{[\ion{O}{3}]}
\def\oi{[\ion{O}{1}]}
\def\nii{[\ion{N}{2}]}
\def\sii{[\ion{S}{2}]}

\begin{document}

\title{Statistical Properties of Radio Emission from the Palomar Seyfert 
Galaxies}

\author{James S.~Ulvestad}
\affil{National Radio Astronomy Observatory}
\affil{P.O. Box O, Socorro, NM 87801; julvesta@nrao.edu}
\and
\author{Luis C.~Ho}
\affil{The Observatories of the Carnegie Institution of Washington}
\affil{813 Santa Barbara St., Pasadena, CA 91011; lho@ociw.edu}

\begin{abstract}

We have carried out an analysis of the radio and optical properties of a 
statistical sample of 45 Seyfert galaxies from the Palomar spectroscopic
survey of nearby galaxies.  We find that the space density of bright 
galaxies ($-22$ mag $\leq M_{B_T} \leq -18$ mag)
showing Seyfert activity is $(1.25\pm 0.38)\times 10^{-3}$~Mpc$^{-3}$,
considerably higher than found in other Seyfert samples.  Host galaxy
types, radio spectra, and radio source sizes are uncorrelated with
Seyfert type, as predicted by the unified schemes for active galaxies.
Approximately half of the detected galaxies have flat or
inverted radio spectra, more than expected based on previous samples.
Surprisingly, Seyfert 1 galaxies are found to have somewhat stronger
radio sources than Seyfert 2 galaxies at 6 and 20~cm, particularly
among the galaxies with the weakest nuclear activity.  We suggest
that this difference can be accommodated in the unified schemes if
a minimum level of Seyfert activity is required for a radio source
to emerge from the vicinity of the active nucleus.  Below this level,
Seyfert radio sources might be suppressed by free-free absorption
associated with the nuclear torus or a compact narrow-line region, 
thus accounting for both the weakness of the radio emission and the 
preponderance of flat spectra.  Alternatively, the flat spectra and
weak radio sources might indicate that the weak active nuclei are
fed by advection-dominated accretion disks.

\end{abstract}


\section{Introduction}
\label{sec:intro}

In the 1970s, significant numbers of Seyfert galaxies were first
identified by means of the Markarian surveys of ultraviolet-excess
objects, the last of which was published by Markarian, Lipovetskij, \&
Stepanian (1981).  Since then, a variety of Seyfert samples
have been developed, including heterogeneous samples derived from
literature searches, and others selected based on specific observational
criteria.  Over the last 25 years, a number of researchers have 
performed arcsecond-resolution interferometric radio surveys of 
the galaxies in these samples.
Among the most important radio surveys or compilations are those 
published by de Bruyn \& Wilson (1976, 1978), Meurs \& Wilson (1984), 
Ulvestad \& Wilson (1984a, 1984b, 1989), Unger et al. (1987), 
Roy et al. (1994), Kukula et al. (1995), Nagar et al. (1999), 
Morganti et al. (1999), Thean et al. (2000, 2001), and Schmitt et al.
(2001a,b).  A comprehensive description of 
the observations, as well as possible selection effects in the different 
samples, is given by Ho \& Ulvestad (2001).

The purposes of studying large statistical samples of Seyfert galaxies
are many.  Such studies can provide important tests of the ubiquity
and applicability of ``unified schemes'' for Seyfert galaxies 
(Antonucci 1993; Wills 1999).  In these schemes, the active nucleus is powered by a 
massive compact object, presumed to be a black hole, surrounded by an 
accretion disk or torus, and many observed properties of the galaxies are 
determined by the observer's line of sight relative to the torus.  The 
Seyfert 1 galaxies, with broad permitted optical emission lines, are those 
in which the torus is viewed closer to face-on orientation.  The Seyfert 2 
galaxies lack broad permitted lines, except (in at least some cases) 
in polarized flux (Antonucci \& Miller 1985; Tran 1995; Moran et al. 2000),
and are thought to be seen with the line of sight 
passing through the torus in an orientation closer to edge-on.  In general, the 
centimeter-wavelength emission from Seyfert nuclei is thought to be 
unobscured, and so should have the same power when the torus is viewed from 
any angle.  However, if that emission is predominantly due to radio jets, those
jets should be viewed closer to end-on in Seyfert 1 galaxies. In this case, 
the nuclear radio sources might tend to have smaller apparent sizes than in 
Seyfert 2 galaxies, although this could be complicated by the fact
that the end-on objects also could have higher surface brightness
and be detected more easily.

Recently, Ho \& Ulvestad (2001) used the Very Large Array (VLA) to image 
52 nearby Seyfert galaxies selected from the Palomar spectroscopic survey of
nearby galaxies (Ho, Filippenko, \& Sargent 1995, 1997a, 1997b; Ho et al. 
1997c), and presented the detailed observational data from that sample.  The 
sample members were chosen from 486 bright northern galaxies typically having 
$B_T \leq 12.5$~mag; the sample is complete to $B_T= 12.0$~mag, and 80\% 
complete to $B_T= 12.5$~mag.  The spectroscopic observations (Ho et al. 1995) 
employed varying integration times, with each target observed long enough to 
reach an approximately  uniform equivalent-width limit of 0.25 \AA\ for 
H$\alpha$ emission (Ho et al. 1997a).  Ho \& Ulvestad (2001) have argued that this
sample was uniformly selected and is expected to be unbiased relative
to many other Seyfert galaxy samples studied in the literature.  In this 
paper, we assess the statistical properties of the radio emission from this
sample, including both the comparisons between Seyfert types and the 
overall correlations with a large set of galaxy properties.

\section{The Radio Data and the Sample}
\label{sec:data}

The radio observations of the Palomar Seyfert sample were described
in detail by Ho \& Ulvestad (2001), but are briefly summarized here for
completeness.  Each of the galaxies in the sample was observed
at two wavelengths, 6 and 20~cm, with the VLA (Thompson et al. 1980).  
The integration time on each galaxy  at each wavelength ranged 
from 15 to 18 minutes, yielding a typical rms noise of 
40~$\mu$Jy~beam$^{-1}$.  Scaled arrays were used, to achieve a maximum
resolution of approximately 1\farcs1 at the two wavelengths; this corresponds 
to a resolution of 110~pc at the typical galaxy distance of $\sim 20$~Mpc.
Full-resolution and tapered images were produced for each galaxy, with the 
tapered images having respective beam sizes of about 2\farcs5 and 3\farcs6.
Radio sizes and position angles of the nuclear sources were measured,
and the flux densities also were found for both the unresolved cores
and any resolved emission associated with the active nucleus.
(Note that extended radio emission apparently associated with the
host galaxy rather than with the active nucleus was not included.)
The results of the radio measurements were tabulated by Ho \& Ulvestad
(2001), together 
with other properties of the host galaxies; those tabulated data form the
primary numerical input to the studies reported in this paper.

The formal definition of the Palomar survey includes all bright 
($B_T \leq 12.5$~mag) northern ($\delta > 0^\circ$) galaxies selected from 
the Revised Shapley-Ames Catalog of Bright Galaxies (RSA; Sandage \& Tammann 
1981).  A total of 52 Seyfert galaxies were imaged with the VLA by 
Ho \& Ulvestad (2001), most of which fit into the two primary selection criteria.  
However, three 
galaxies (NGC~1068, NGC~1358, and NGC~1667) had negative declinations, so they
should be excluded from the statistical sample.  In addition, a few
other Seyferts fainter than the formal magnitude limit were imaged.  These
galaxies were included in the original sample either because of historical 
interest or because their apparent magnitudes had been revised 
slightly since the 
RSA was published (Ho et al. 1995).  For our statistical sample, we have 
chosen to eliminate four galaxies with current measurements
of $B_T > 12.64$ mag: NGC~1167, NGC~3185, NGC~4169, and NGC~5548.  At
brighter magnitudes, the completeness corrections for the RSA are
relatively minor and well established (Sandage, Tammann, \& Yahil 1979).  
Therefore,
the final statistical sample of Palomar Seyferts, studied in the rest
of this paper, includes a total of 45~galaxies.  Of these,
20 are type~1 Seyferts (defined to include all those types showing
detectable broad-line emission, from type 1.0 through 1.9), and 25
are type~2 Seyferts.

\section{Statistical Properties and Correlations with Seyfert Type}
\label{sec:stat}

A large number of properties of the Seyfert radio sources have been studied 
statistically.  The statistical studies made use of an array of tests for 
censored data based on the package ASURV described by Isobe \& Feigelson (1990) 
and by LaValley, Isobe, \& Feigelson (1992).  
We generally use Gehan's Generalized Wilcoxon Test (hereafter 
``Gehan-Wilcoxon test'') to assess possible differences between two samples. 
This test is used to define the probability that two
samples drawn from the same parent population would have a difference as
large as that observed in a particular parameter.
The generalized Kendall's $\tau$ coefficient is used to assess correlations
between two variables.
Distance-dependent quantities in this paper are based 
on a Hubble constant of $H_0 = 75$~km~s$^{-1}$~Mpc$^{-1}$; when necessary, 
comparison samples from the literature have been converted to this distance 
scale.  In the subsections below, we present the results of various tests and
some comparisons to previously published results for Seyfert samples.

\subsection{Galaxy Distances}
\label{sec:dist}

The median distance of the parent sample of Palomar galaxies is 18.2~Mpc 
(Ho et al. 1997a).  Histograms of the distances of the galaxies classified 
as Seyferts are given in
Figure~1; the median distances are 17.0~Mpc for the type 1
Seyferts and 20.4~Mpc for the type 2 Seyferts.  These typical distances are
considerably smaller than for most samples of Seyfert galaxies
studied at radio wavelengths.  For example, the distance-limited
sample studied by Ulvestad \& Wilson (1989) has a median distance of approximately 
35~Mpc, while the CfA sample (Huchra \& Burg 1992) has a median distance of 80~Mpc.
The Gehan-Wilcoxon test shows that the distances of the two Seyfert types in 
the Palomar sample would differ by the observed amount 54\% of the time
if they were drawn from the same parent sample;
thus, there are no significant distance 
biases between the
two Seyfert types, which conceivably could cause systematic effects in 
quantities such as their radio luminosities.

\vskip 0.3cm

\psfig{file=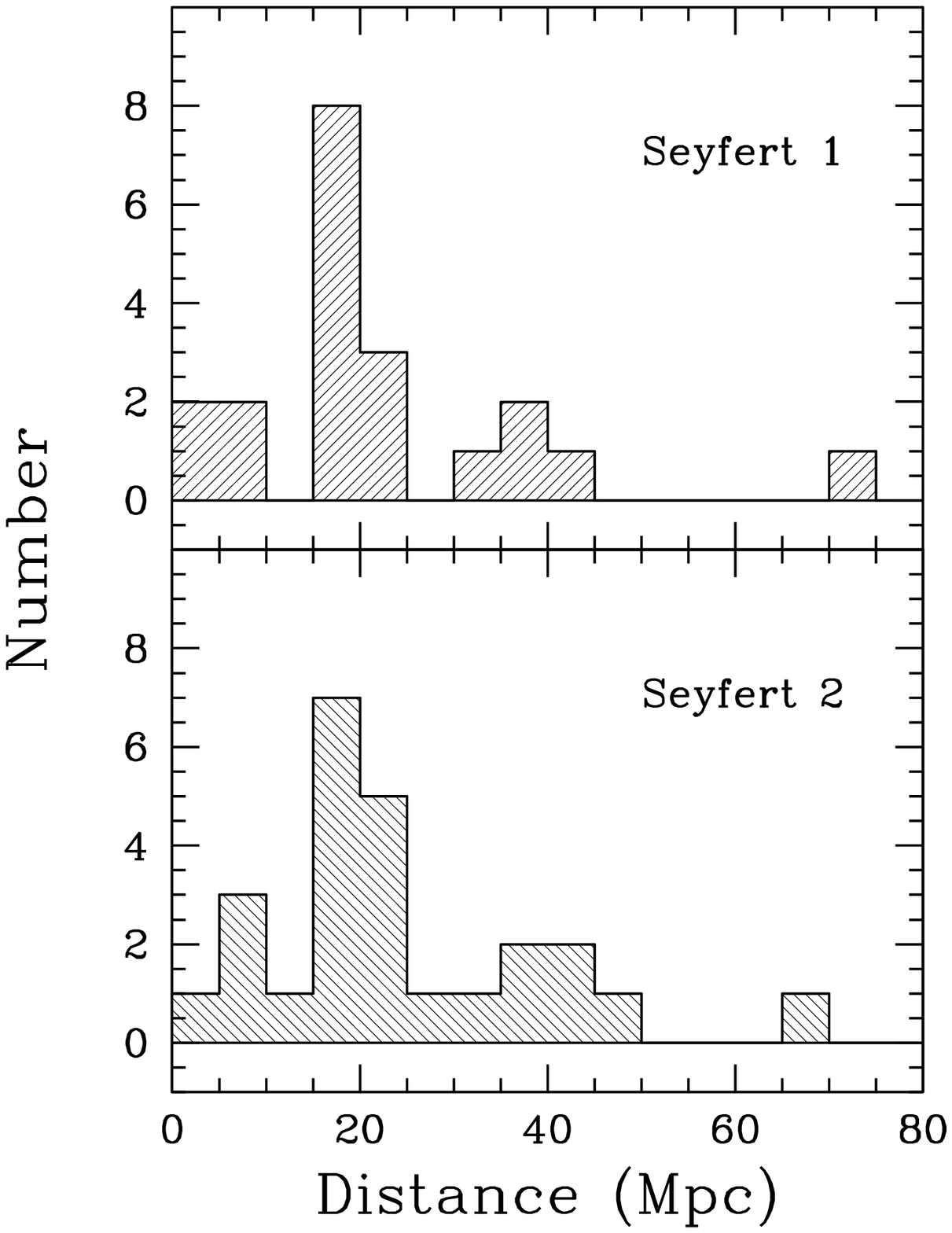,width=8.5cm,angle=0}
\figcaption[f1.eps]{
Histogram of galaxy distances in the Palomar
Seyfert sample.}
\label{fig:dist}
\vskip 0.3cm


\subsection{Host-Galaxy Types}
\label{sec:host}

Ho \& Ulvestad (2001) gave host-galaxy morphological types for all the objects in the 
Palomar Seyfert sample.  After removing the peculiar galaxy NGC~1275, whose 
morphology is ambiguous, the remaining 44 objects have been tested to see if 
there is any difference between the host-galaxy types of the type 1 and 
type 2 Seyferts.  The Gehan-Wilcoxon test shows a 48\% probability that the 
host galaxy types of the type 1 and type 2 Seyferts would differ as much
as observed if they were drawn from 
the same parent sample, so there is no significant correlation
between galaxy type and Seyfert class.  As noted by Ho et al. (1997b), 4 of
the 45 Palomar Seyferts reside in elliptical host galaxies.
Often Seyfert galaxies are defined to be in spiral hosts, but this is
not a requirement of the definition of Seyfert galaxies used for
the Palomar sample, in spite of the classical ``lore'' that all 
Seyferts are in spiral galaxies.

\subsection{Optical Luminosities}
\label{sec:optlum}

The apparent and absolute magnitudes of the Palomar Seyferts have been
tabulated by Ho \& Ulvestad (2001), and can be used to derive an optical luminosity
function.  To do so, we have used the $V/V_{\rm max}$ method 
(Schmidt 1968; Huchra \& Sargent 1973; Condon 1989),
correcting for the incompleteness of the RSA catalog as specified by Sandage 
et al. (1979); the maximum completeness correction is a factor of 1.5.  Errors
have been estimated using the method described by Condon (1989).
Table~1 gives the optical luminosity function as a function of 
absolute magnitude corrected for Galactic extinction, $M_{B_T}$, as given in 
Ho et al. (1997a).  Note that Ho \& Ulvestad (2001) tabulated absolute magnitudes
that were corrected for both Galactic {\it and}\ internal extinction, 
$M_{B_T}^0$; here, we do not apply the correction for internal extinction, in 
order to facilitate comparison with other samples.  Figure~2 
compares our luminosity function to the result for the CfA Seyferts,
adapted from Huchra \& Burg (1992). The optical magnitudes of the CfA Seyferts were 
not corrected for Galactic extinction, but this correction is expected to be 
small at the high Galactic latitudes of that sample.  
We stress that the optical luminosity functions 
pertain to the {\it integrated}\ light of the whole galaxy (nucleus plus 
host), rather than just the active nucleus alone, which can be significantly 
fainter (Ho \& Peng 2001).

\vskip 0.3cm

\psfig{file=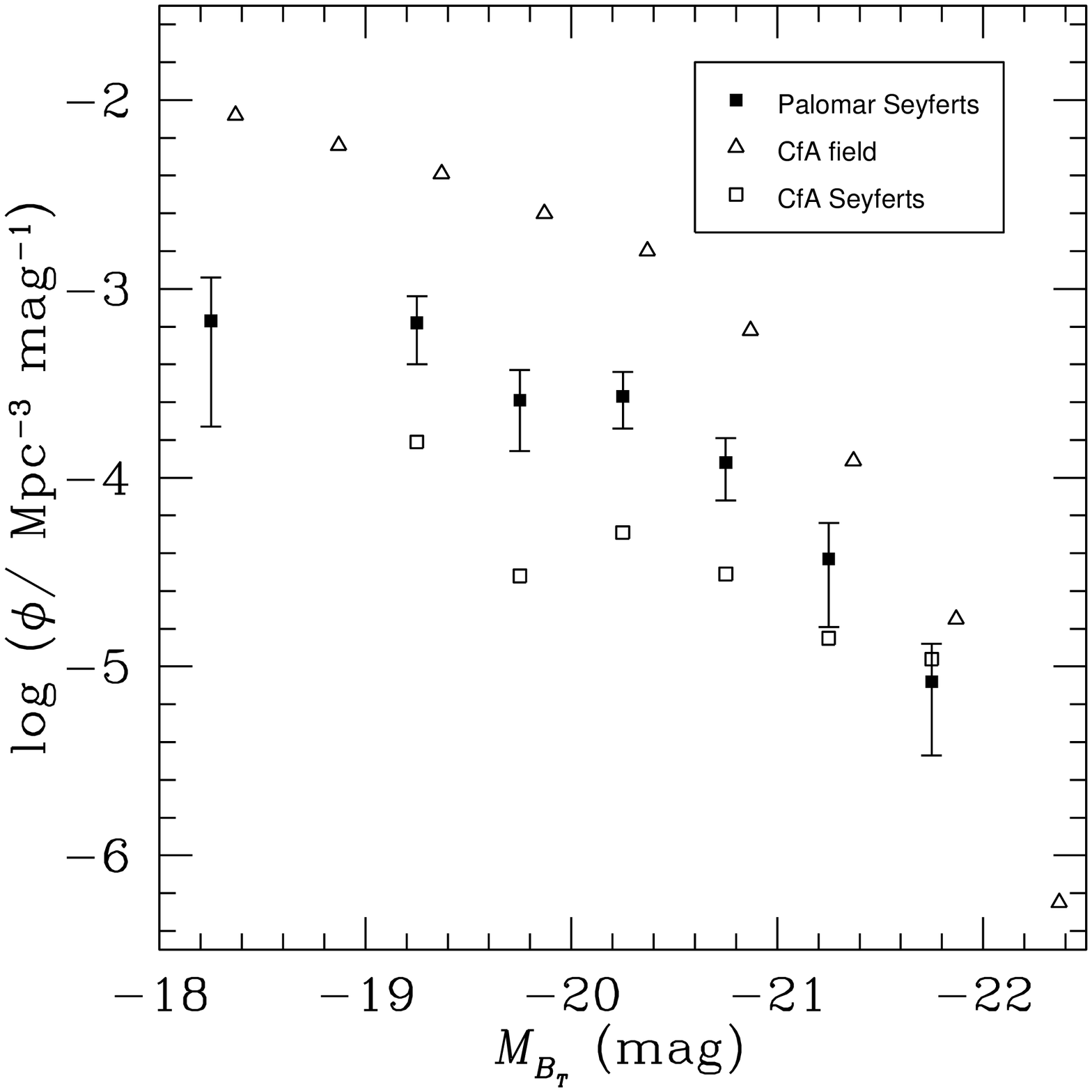,width=8.5cm,angle=0}
\figcaption[f2.eps]{
Optical luminosity function of Palomar Seyferts,
compared to CfA Seyferts and CfA field galaxies.  The
CfA data are from Huchra \& Burg (1992), converted to
$H_0=75$~km~s$^{-1}$~Mpc$^{-1}$.  To reduce clutter,
error bars have been plotted only for the Palomar Seyferts.
Only bins containing more than one galaxy have been plotted.}
\label{fig:optlum}
\vskip 0.3cm

From the data shown in Figure~2, it is evident that the
derived luminosity function for the Palomar Seyferts is a factor of
5--10 higher than the CfA Seyferts for $-22$ mag $\leq M_{B_T} \leq -19$ mag.  This is 
consistent with the fact that of the 35 galaxies in the statistical Palomar 
Seyfert sample that are within the selection limits of the CfA 
survey, only 10 were recognized as Seyferts in that survey.
The derived luminosity function also shows that the space density of Palomar 
Seyferts may approach that of the field galaxies in the CfA sample
near $M_{B_T} \approx -22$ mag, implying that a large fraction of 
galaxies of this total luminosity host at least a weak Seyfert nucleus.

\subsection{Radio Luminosities}
\label{sec:radlum}

\subsubsection{Seyfert 1 vs. Seyfert 2 Galaxies}
\label{sec:radcomp}

Over the years, many studies have been made of the radio luminosities of 
Seyfert galaxies, and the relation to Seyfert type.  Early studies generally 
indicated that Seyfert 2 galaxies are more luminous radio sources than 
Seyfert 1 galaxies at centimeter wavelengths for objects obeying the
same (sometimes ill-defined) selection criteria (de Bruyn \& Wilson 1978;
Ulvestad \& Wilson 1984a,b). However, the ultraviolet selection for
Markarian and other Seyfert galaxies strongly biased the early samples
toward Seyfert 2 galaxies that were much higher on the radio luminosity 
function than the Seyfert 1 galaxies (see discussion in Ho \& Ulvestad 2001, 
\S~A.2).  Later papers (Edelson 1987; Ulvestad \& Wilson 1989; 
Giuricin et al. 1990) seemed to indicate that the 
discrepancy between the two Seyfert types disappeared with less biased samples,
although these studies also suffered from poor resolution (Edelson 1987) and 
from sample heterogeneity (Ulvestad \& Wilson 1989; Giuricin et al. 1990).  
Most recently, high-resolution 
3.6~cm VLA imaging has shown that Seyfert 1 and Seyfert 2 galaxies have 
indistinguishable radio luminosities in samples selected at 12~$\mu$m 
(Thean et al. 2001) and at 60~$\mu$m (Schmitt et al. 2001a).  Therefore, 
a general consensus
has evolved that Seyfert 1 and Seyfert 2 
galaxies have statistically similar radio luminosities, a result which has 
served as a strong argument in favor of the unified schemes for Seyferts.

Ho \& Ulvestad (2001) provided measures of the 6 and 20~cm radio luminosities
of the Palomar Seyferts, both in cores unresolved by the VLA and in
the total emission associated with the active nuclei.  Histograms of
the total radio luminosities are shown in Figure~3; 
these differ slightly from the plots shown by Ho \& Ulvestad (2001), 
because we consider here the more restricted 
sample of 45 Seyferts.  The type 1 Seyferts appear to be somewhat more 
luminous in the radio than the type 2 Seyferts, exactly the opposite of
the early studies that showed apparent differences between the Seyfert types.
For example, at 6~cm, all eight undetected objects are Seyfert 2
galaxies; at 20~cm, 13 of the 14 undetected objects are type 2 Seyferts.
More quantitatively, the Gehan-Wilcoxon test gives a 2.0\% probability that 
the 6~cm core powers of the type 1 and type 2 Seyferts would differ as much
as observed if they were drawn from
the same parent distribution; this probability is 1.3\% for
the total 6~cm emission of the two Seyfert types. 
At 20~cm, both the core and total radio
powers appear to be correlated with Seyfert type, with a probability
of only 0.9\% that the Seyfert 1 and Seyfert 2 galaxies would look as
different as observed if they were drawn
from the same parent population.  

\vskip 0.3cm

\psfig{file=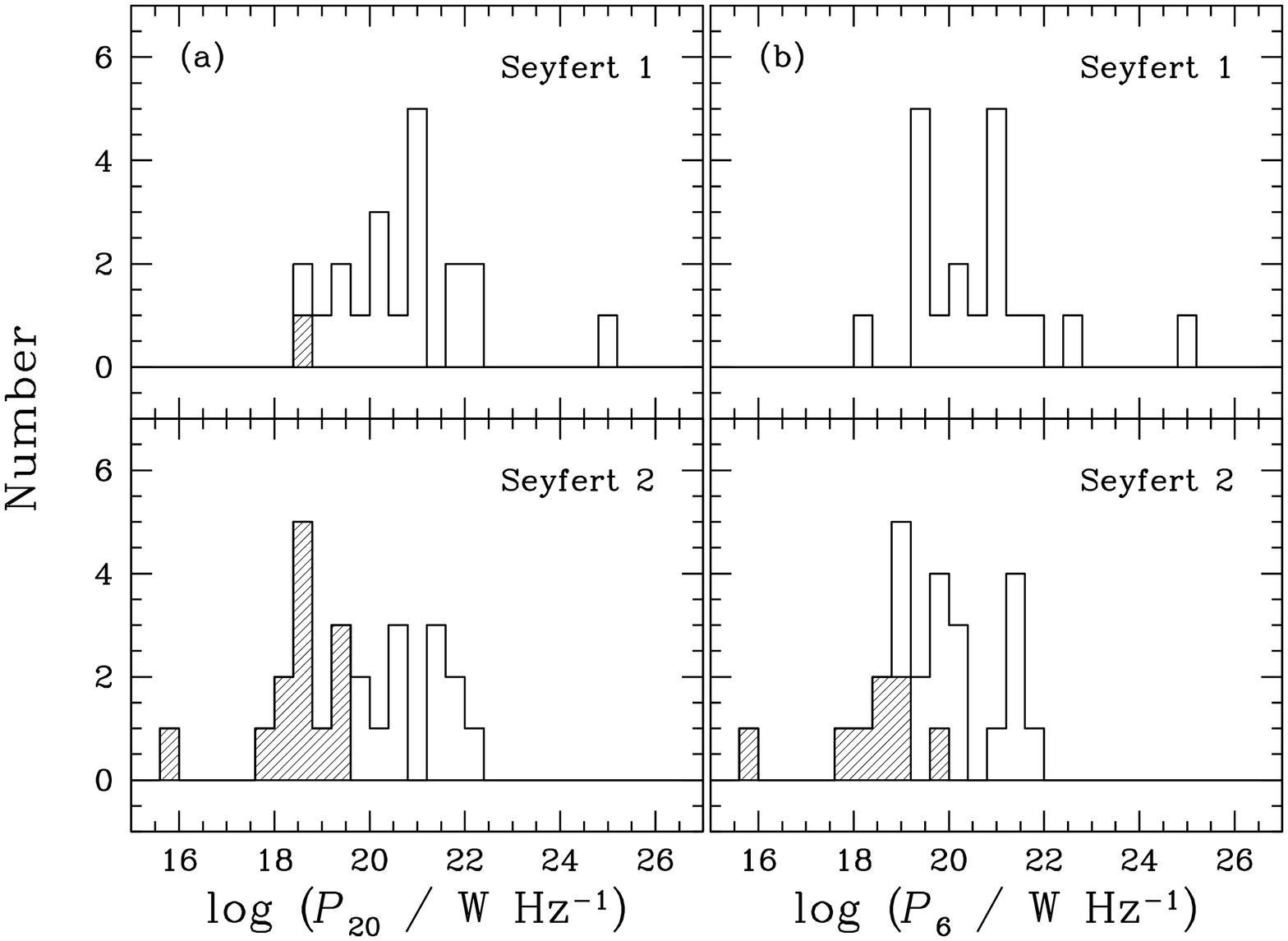,width=9.5cm,angle=270}
\vspace{-1.0cm}
\figcaption[f3.ps]{
Histogram of ({\it a\/}) the total 20~cm radio powers and 
({\it b\/}) the total 6~cm radio powers for the galaxies in the Palomar Seyfert 
sample.  Upper limits are shown hatched.}  
\label{fig:powerhist}
\vskip 0.3cm

These results would appear to swing the pendulum back in the direction of 
there being intrinsic differences between the radio luminosities of Seyfert 1 
and Seyfert 2 galaxies, with the Seyfert 1 galaxies being stronger rather than 
weaker.  Of course, any bias toward more powerful Seyfert 1 galaxies being 
selected from the Palomar galaxy survey also could account for the apparent 
differences.  The effect is not due to distance; as noted in \S~3.1, the 
Seyfert~1 galaxies are, in fact, slightly closer (on average) than the 
Seyfert 2 galaxies in our sample.  Further discussion on this unexpected 
result is deferred to \S~4.3 and 4.4.

\subsubsection{Radio Luminosity Functions}
\label{sec:radlf}

Radio luminosity functions can be derived for our optically 
selected sample by first finding the bivariate optical-radio luminosity 
function, and then using that to derive the radio luminosity function, as
was done by Meurs \& Wilson (1984).  We have carried out this procedure for both the
6 and 20~cm observations, computing the luminosity functions down to values 
below which more than 75\% of the measurements are upper limits.
Extending the analysis to lower luminosities would be subject
to large uncertainties, due to the fact that very few detections are 
present in the same luminosity range as most of the upper limits.
At 6~cm, where the upper limit for NGC~5395 fell well 
above many detections, this limit has been distributed among the lower bins 
according to the population of sources in those bins.  However, since 
$V_{\rm max}$ is quite high for NGC~5395, this galaxy has relatively low 
weight and does not affect the results appreciably.

Table~2 lists both the 6 and 20~cm luminosity
functions. Figure~4{\it a}\ shows the 20~cm luminosity
function compared to results from other samples (most of which were
derived at 20 cm), while the 6~cm luminosity function for the 
Palomar Seyferts is in Figure~4{\it b}.
We make no effort to distinguish between the two Seyfert types, because each 
of the samples contains $\lesssim 50$ galaxies, insufficient to populate 
most of the luminosity bins adequately for each type.  
In Figure~4{\it a}, the 20~cm luminosity functions found by 
Meurs \& Wilson (1984) for the Markarian Seyferts and
by Ulvestad \& Wilson (1989) for a distance-limited Seyfert sample have
been converted to our adopted value of $H_0 = 75$~km~s$^{-1}$~Mpc$^{-1}$.

\vskip 0.0cm

\psfig{file=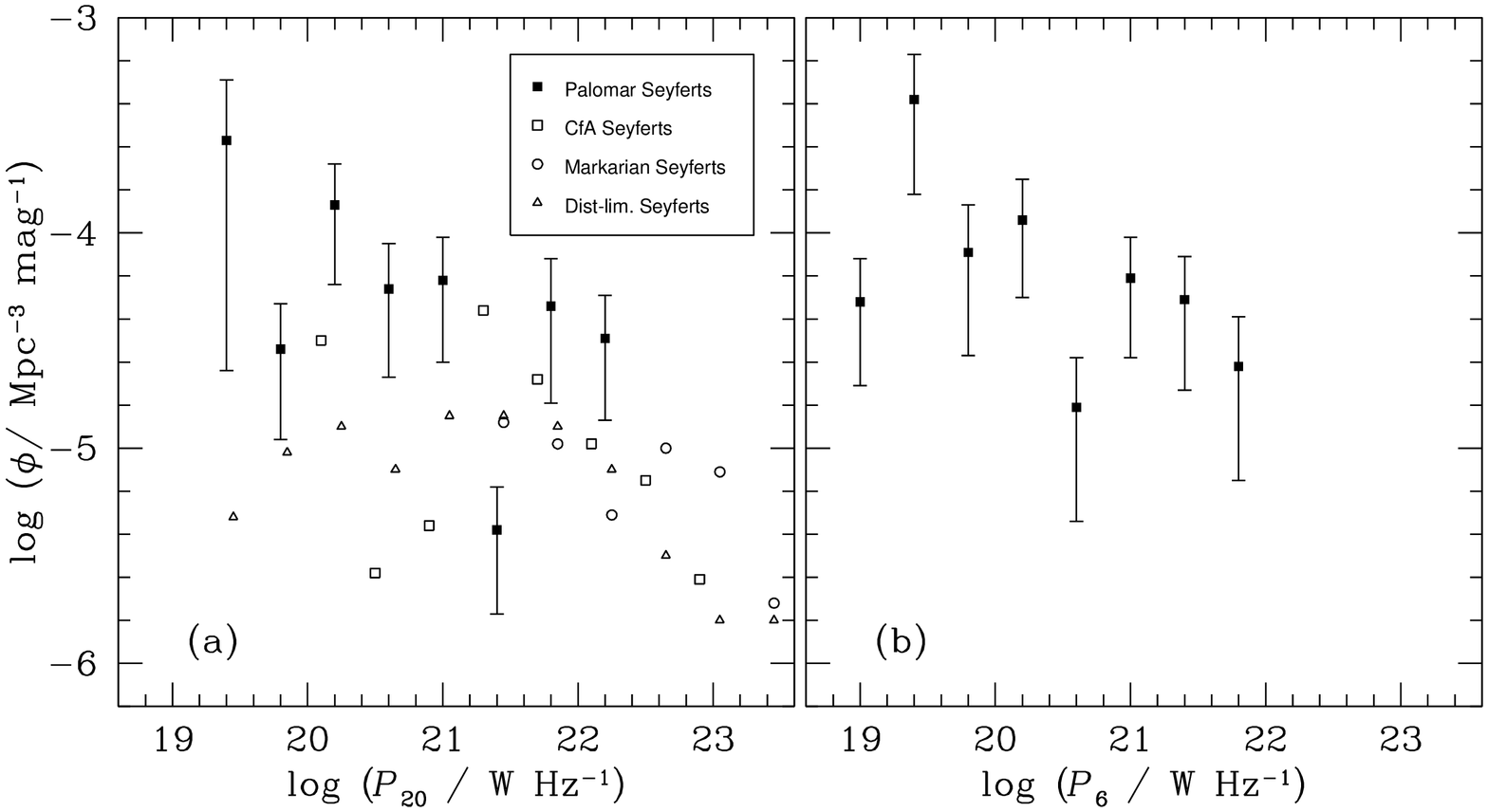,width=9.5cm,angle=270}
\vspace{-1.5cm}
\figcaption[f4.ps]{
Radio luminosity functions of Palomar Seyferts at ({\it a\/}) 20~cm 
and ({\it b\/}) 6~cm.  Only bins containing more than one galaxy 
have been plotted.  Panel ({\it a\/}) also shows data for the Markarian Seyferts 
(Meurs \& Wilson 1984), distance-limited Seyferts (Ulvestad \& Wilson 1984b,
1989) and CfA Seyferts (Kukula et al. 1995).  The luminosities of the CfA 
Seyferts have been converted from 3.6~cm to 20~cm using a spectral index of 
$\alpha=-0.7$, and their points are offset to the left by 0.1 in $\log P_{20}$ for 
clarity.}
\label{fig:radlum}
\vskip 0.3cm

Kukula et al. (1995) did not present a radio luminosity function for the CfA 
Seyferts.  We have converted their C-configuration 3.6~cm flux densities to 
20~cm using a spectral index of $-0.7$, and then computed the luminosity
function using the $V/V_{\rm max}$ technique.  We used their 
low-resolution (2\arcsec ) flux densities rather than the values at 0\farcs25 
resolution because the angular scale sampled is more comparable to 
the total flux densities used for the Palomar Seyfert sample.
As a check, we compared the 20~cm flux densities estimated in
this way to those measured by us for nine galaxies in common between
the CfA and Palomar samples.  The mean difference is
$\langle\log(P_{\rm CfA}) - \log(P_{\rm Pal})\rangle\, =\, 0.09$~dex, with 
an rms scatter of 0.25~dex.  Except for NGC~4235, which actually is
known to have a much flatter spectrum than the assumed value
of $-0.7$ (Ulvestad \& Wilson 1989), the remaining eight galaxies all have
20~cm powers extrapolated from the Kukula et al. (1995) measurements that are
within 0.25 dex of those measured directly by Ho \& Ulvestad (2001).

Figure~4{\it a}\ shows that the radio luminosity functions for the Palomar, 
CfA, distance-limited, and Markarian samples are consistent with
one another for $P_{20}\,\gtrsim\,10^{21}$~W~Hz$^{-1}$, given the small numbers 
of galaxies present.  It appears that the Palomar Seyferts have significantly  
higher space densities at radio luminosities below $\sim 10^{21}$~W~Hz$^{-1}$.
This is consistent with expectations, because the Palomar sample generally 
includes more nearby and weaker Seyferts than the other three samples, which 
are undoubtedly incomplete at the lower power levels.  Indeed, the luminosity 
function for the Markarian sample does not even extend to values below 
$P_{20}\,\approx\,10^{21.5}$~W~Hz$^{-1}$.


\begin{figure*}[t]
\vspace{-2.5cm}
\centerline{\psfig{file=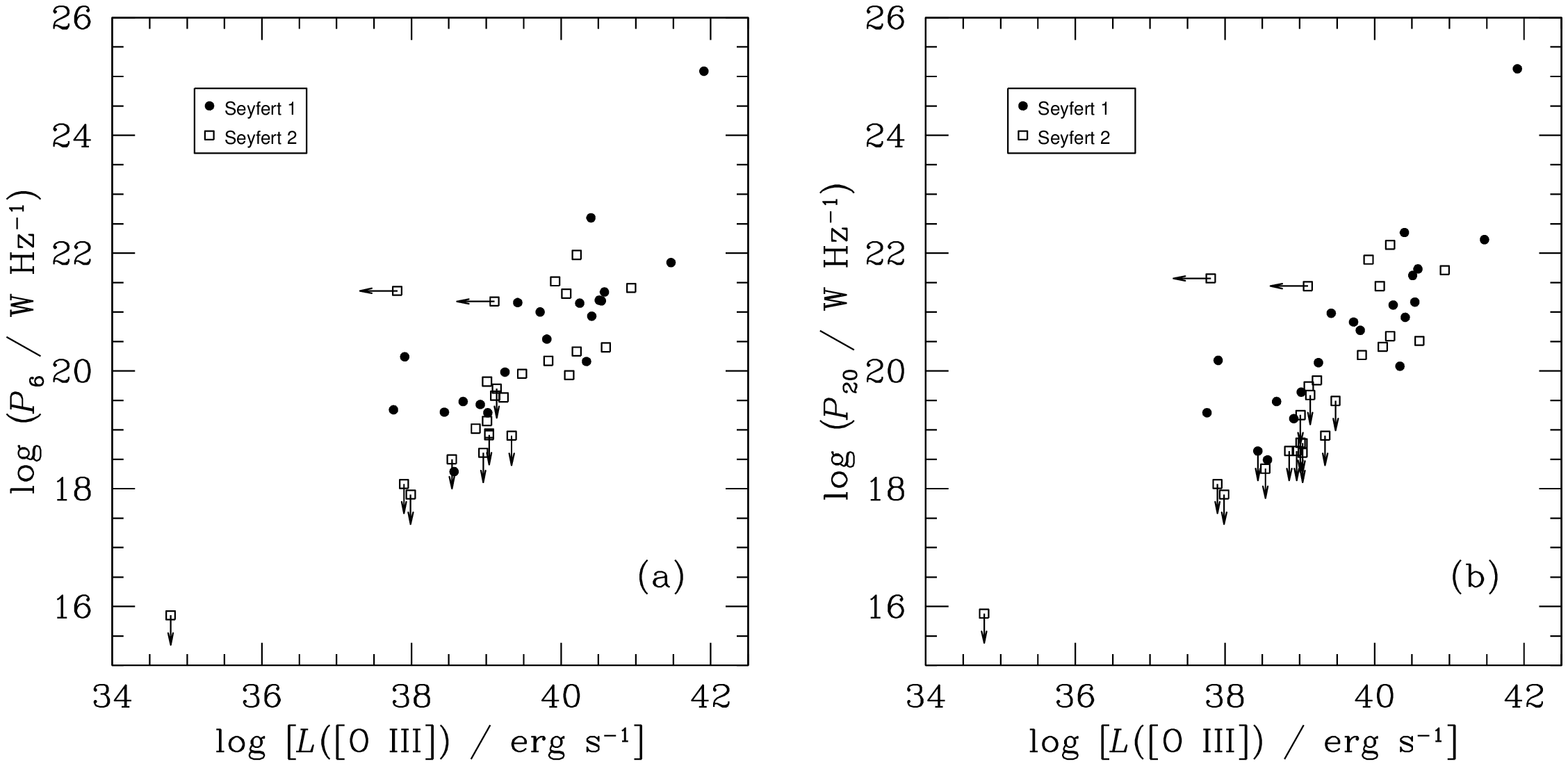,width=19.5cm,angle=270}}
\vspace{-3.2cm}
\figcaption[f5.ps]{
Plots of ({\it a\/}) 6~cm radio power and ({\it b\/}) 20~cm radio 
power {\it vs.} the extinction-corrected \oiii\ luminosity for Palomar Seyferts.  
Seyfert 1 and 2 galaxies are distinguished by closed and open symbols,
respectively.
\label{fig:rad-oiii}}
\end{figure*}
\vskip 0.3cm

\subsection{Correlations between Radio and Optical Emission Lines}
\label{sec:oiii}

Strong correlations of 20~cm continuum power with \oiii\ $\lambda$5007 
emission-line luminosity and line width have been well established for 
Seyferts (de Bruyn \& Wilson 1978; Wilson \& Willis 1980; Heckman et al.
1981; Meurs \& Wilson 1984; Whittle 1985); the basic results were
summarized by Wilson (1991).  Whittle (1985) showed that the relationship 
between radio power and line luminosity appears to hold over several orders 
of magnitude in each variable. Whittle (1992a, 1992b) later showed that the 
relation between radio luminosity and \oiii\ line width indicates anomalously 
large widths in powerful linear radio sources ($P_{20} >
10^{22}$--$10^{23}$~W~Hz$^{-1}$), which appear to
accelerate narrow-line gas beyond the values expected for
virialized motions.  However, the correlation still held for
galaxies with weaker radio sources, indicating a remaining 
relationship with bulge luminosity.  For Seyfert~1 galaxies,
radio-quiet and radio-loud objects cover the same range of
\oiii\ luminosities, but Ho \& Peng (2001) have shown that 
the correlation 
of radio power with \oiii\ line luminosity actually has two distinct 
relations, one for radio-quiet and one 
for radio-loud objects (where ``radio-loud'' and ``radio-quiet'' are 
defined by using the radio-to-optical luminosity ratio of the active 
nucleus, rather than of the entire galaxy).  

Not surprisingly, computation of the generalized Kendall's $\tau$ coefficient 
for the Palomar Seyfert sample shows that both the 20 and 6~cm radio powers 
are correlated with the \oiii\ luminosity\footnote{In this and subsequent 
discussions, the \oiii\ luminosity pertains to the $\lambda$5007 line, 
corrected for Galactic and internal extinction, as described in Ho et al. 
(1997a).}, with the probability of a null correlation being less than 0.01\%.  
Plots of these relationships are shown in Figure~5. 
A similar relationship is obtained by plotting the radio fluxes against
emission-line fluxes (not shown, to save space), so the apparent 
correlation is not just a distance effect.  The actual values of radio 
power at a given \oiii\ strength are scattered over more than two orders of 
magnitude.  For example, at 6~cm, Figure~5{\it a}\ 
shows that the radio power
varies from $~\sim 10^{20}$ to $~\sim 10^{22}$~W~Hz$^{-1}$
at an \oiii\ line luminosity near $10^{40}$~erg~~s$^{-1}$.  Testing the
correlation of \oiii\ luminosity alone against Seyfert type, we find
a 15\% probability that the different Seyfert types would look as different
as observed if drawn from the 
same parent population in \oiii\ luminosity.

Previously, there were indications from the heterogeneous sample of objects 
studied by Whittle (1985) that Seyfert 2 galaxies have higher radio powers 
(by a factor of 5--10) than Seyfert 1 galaxies at a given \oiii\ luminosity.  
We can test to see whether the ratio of radio power to \oiii\ luminosity is 
the same for the two Seyfert types.  In this case, the generalized Kendall's 
$\tau$ coefficient shows that the ratio of total 20~cm power to \oiii\ 
luminosity is significantly different between Seyfert 1 and Seyfert 2 
galaxies; if the two Seyfert types were drawn from the same parent population,
the observed difference would be expected only 0.9\% of the time.
Using the 6~cm power instead, the difference is significant at the 
level of 1.2\%.  The apparent correlation is in the 
sense that, for a given \oiii\ luminosity, the radio power is {\it higher}
for Seyfert~1 galaxies than for Seyfert~2 galaxies, opposite to
the sense suggested by Whittle (1985).

\vskip 0.0cm

\begin{figure*}[t]
\vspace{-2.5cm}
\centerline{\psfig{file=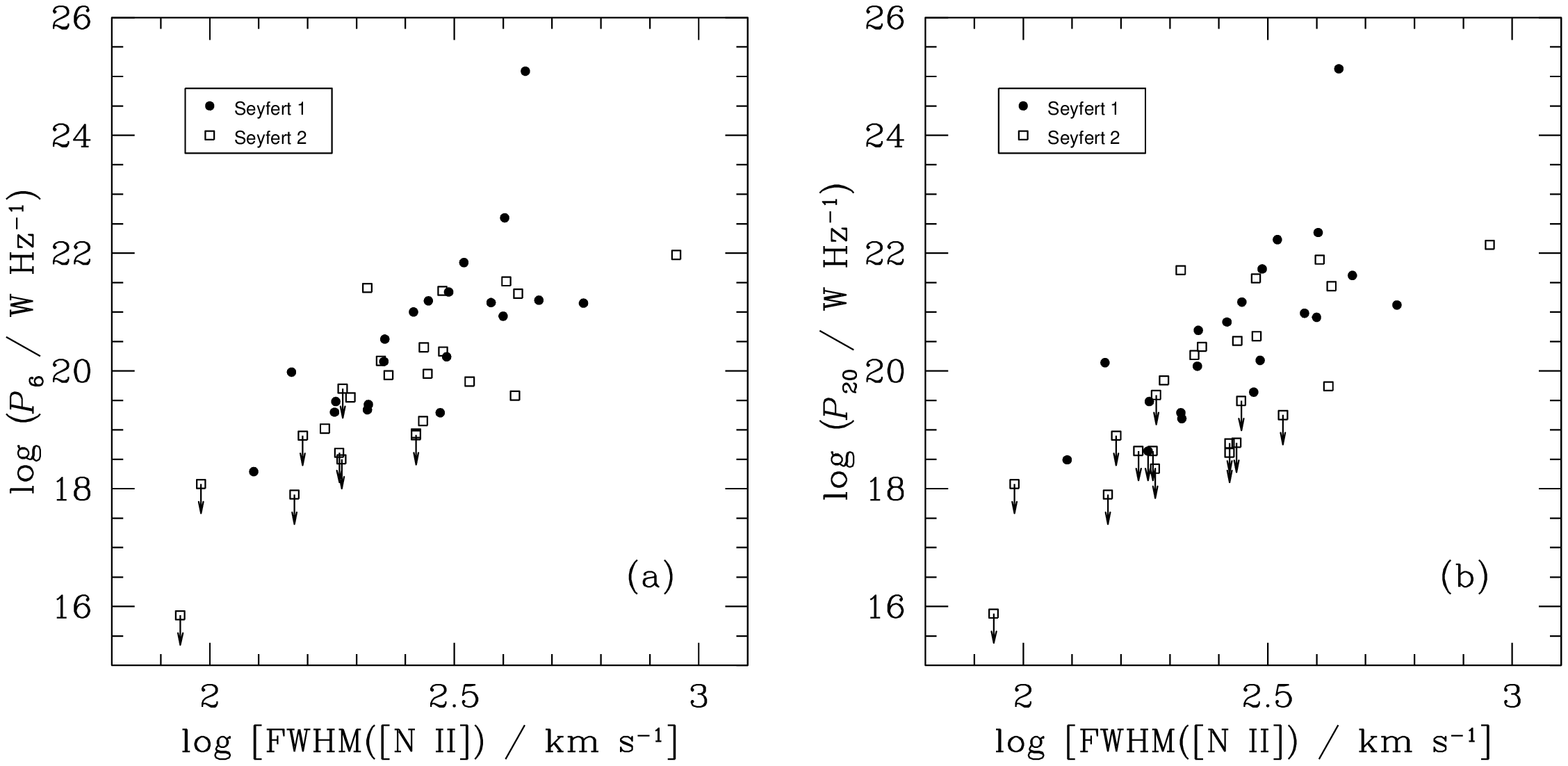,width=19.5cm,angle=270}}
\vspace{-3.3cm}
\figcaption[f6.ps]{
Plots of ({\it a\/}) 6~cm radio power and ({\it b\/}) 20~cm radio 
power {\it vs.} \nii\ line width for Palomar Seyferts.  Seyfert 1 and 2 galaxies 
are distinguished by closed and open symbols, respectively.
\label{fig:rad-nii}}
\end{figure*}
\vskip 0.3cm

Along with the tests of radio vs. optical emission-line strength, we
have tested for correlation of the radio powers at both 6 and 20~cm with the
kinematics of the narrow-line region.  This test traditionally makes use of
the \oiii\ $\lambda$5007 line (Wilson \& Willis 1980; Whittle 1985, 1992a).
For the Palomar survey, however, the red spectral region has higher dispersion
than the blue spectral region, and Ho et al. (1997a) use the 
\nii\ $\lambda$6584 line as a surrogate for \oiii\ $\lambda$5007 to quantify the
width of the narrow emission lines.  In the narrow-line regions of Seyfert 
galaxies, \nii\ and \oiii\ trace roughly similar kinematics (e.g., Busko \& 
Steiner 1992).  We use the values of FWHM(\nii) as given
in Ho \& Ulvestad (2001), but omit NGC~777, for which \nii\ was not detected.  
In this case, we again find a highly significant correlation, with the 
probability of no correlation being less than 0.01\%  for both radio 
frequencies.  Scatter plots are shown in Figure~6.  As for the 
plots of radio power vs. \oiii\ luminosity, a scatter of two orders of 
magnitude or more in the radio power is seen for a given value of the \nii\ 
line width.  Few of the Palomar Seyferts have 20~cm radio powers above 
$10^{22}$~W~Hz$^{-1}$, so we would expect few of the outliers with
high line width that were identified in strong linear radio sources by 
Whittle (1992a).  In fact, the one galaxy with a strong radio power and an
anomalously high line width is NGC~3079 ($\log P_{6}$ = 21.97, 
$\log P_{20}$ = 22.14, FWHM(\nii) = 900 km~s$^{-1}$), which appears to be
dominated by a diffuse radio source rather than a linear source
(Ho \& Ulvestad 2001).

\subsection{Radio Spectra}
\label{sec:spectra}

Most previous Seyfert samples observed at radio wavelengths do not
have high-resolution, multi-frequency spectral information.  An
exception is the distance-limited sample of Ulvestad \& Wilson 
(1984b, 1989), which includes arcsecond imaging at 6 and 20~cm for a
total of 43 galaxies.  Only two of those 43 galaxies have ``flat''
radio spectra, where ``flat'' spectra are defined to have
spectral indices $\alpha\geq -0.2$
(using the definition that $S_\nu\propto \nu^{+\alpha}$).
The distribution of spectral indices from this distance-limited
sample is shown in Figure~7.
However, a cautionary note is that the distance-limited
sample was observed at higher resolution at 6~cm than at 20~cm,
so there could be a tendency for the total fluxes at 6~cm
to be underestimated relative to 20~cm. This might yield steeper spectra than 
observations with matched beam sizes, as was achieved with the Palomar sample.

\vskip 0.3cm

\psfig{file=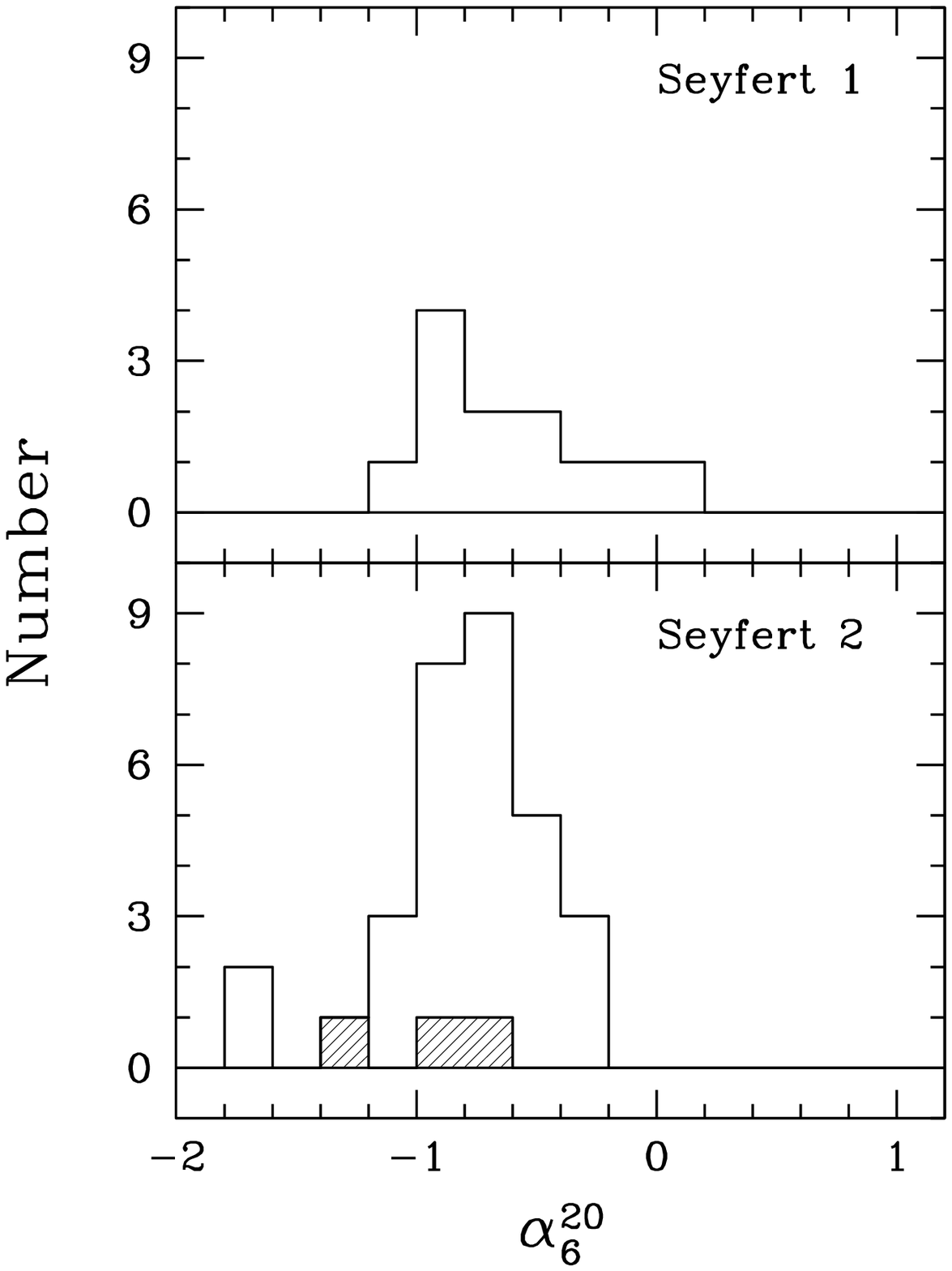,width=8.5cm,angle=0}
\figcaption[f7.eps]{
Histogram of two-point spectral indices
between 20 and 6 cm for the distance-limited
Seyfert sample of Ulvestad \& Wilson (1984b, 1989).
Upper limits (detections at 20~cm and upper limits at 6~cm)
are indicated by hatched regions.}
\label{fig:dlspect}

\vskip 0.3cm

Ho \& Ulvestad (2001) gave two-point radio spectral indices for the galaxies in
the Palomar Seyfert sample that were detected at both 6 and
20~cm.  We have added to these values the limits for the objects that
were detected only at 6~cm.  In addition, in order to isolate any core 
component which may potentially lie within the region of the torus,
we have computed the spectral indices from the peak emission at the
highest resolution, with similar ({\it u,v}) coverage and
identical restoring beams at 6 and 20~cm.  
Figure~8 shows histograms of the spectral
indices of the peak and total radio emission for the
37 galaxies in the statistical sample that were detected
at 6~cm and/or 20~cm.  Fifteen of these galaxies have
total spectral indices of $\alpha \geq -0.2$, while 19
have spectral indices of their peaks falling in the same range.

\vskip 0.3cm

\psfig{file=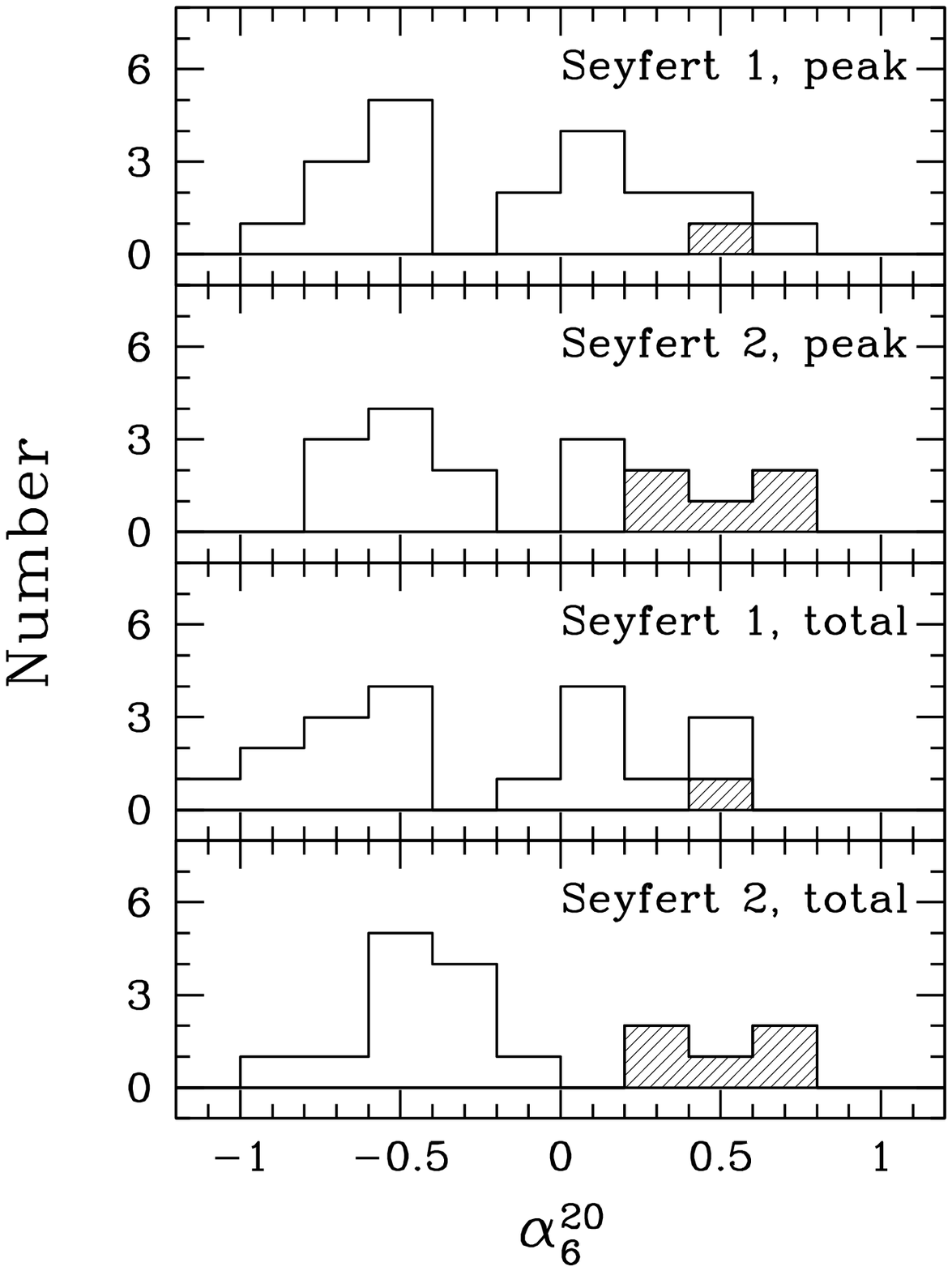,width=9.5cm,angle=0}
\figcaption[f8.eps]{
Histogram of two-point spectral indices
between 20 and 6 cm for the Palomar Seyfert galaxies.
Lower limits (detections at 6~cm and upper limits at 20~cm)
are indicated by hatched regions.}
\label{fig:spect}

\vskip 0.3cm

There is a clear difference between the Palomar sample and 
the distance-limited sample of Ulvestad \& Wilson (1984b, 1989), 
with a probability of less than 0.01\% that the
two distributions would differ by the observed amount if 
drawn from the same parent population (even though
a few galaxies are in both samples).  
Given the large number of objects in both samples that are 
unresolved or slightly resolved, we do not expect possible
resolution effects in the distance-limited sample to
account for the dramatic differences in spectral indices.  For 
example, if we consider only those sources that were unresolved in
the distance-limited sample, only 18\% (2 of 11) have 
$\alpha\geq -0.2$, substantially fewer than the 41\% (15 of 37)
Palomar Seyferts with similarly flat spectra.

There is a slight indication from Figure~8
that Seyfert~2 galaxies have systematically larger
(i.e., flatter or more inverted) radio spectral indices than 
Seyfert~1 galaxies.  Certainly, most of the sources with 
lower limits to their spectral indices are Seyfert 2 galaxies.  
However, the differences are not
statistically significant.  The Gehan-Wilcoxon test gives
a 29\% probability that the total spectral indices of the two
Seyfert types would differ by the observed amount if 
drawn from the same parent population, or a 63\%
probability using the spectral indices of the peaks.

\subsection{Radio Source Sizes}
\label{sec:radsize}

Early suggestions were that Seyfert 2 galaxies were not only more
powerful than Seyfert 1 galaxies at centimeter wavelengths, but also
had larger radio sources (Ulvestad \& Wilson 1984a,b).  After controlling
for the strengths of the radio sources by measuring sizes at a
uniform fraction of the radio peak, Ulvestad \& Wilson (1989) found that this
difference was not statistically significant.  However, in the unified 
schemes, one would expect the Seyfert 1 galaxies to have generally smaller 
radio sources. Since Seyfert 1 galaxies would have more nearly face-on
tori, their radio jets would be expected to lie along the line of sight 
to the observer, and hence be foreshortened.  Indeed, Schmitt et al. (2001a)
have found just this effect in the de Grijp et al. (1992) sample of Seyfert
galaxies selected at 60~$\mu$m.

Figure~9 shows the distribution of radio sizes for the
active-nucleus components of the Seyfert galaxies in the Palomar sample.
Making a statistical study of the radio sizes of these galaxies 
is subject to considerable uncertainty; some radio sources are
unresolved, while others are undetected.  Therefore, we have performed the 
Gehan-Wilcoxon test on two different subsamples of Seyfert galaxies, 
those with measured sizes only (19 galaxies), and those with either measured
sizes or upper limits (37 galaxies).  These tests show respective
probabilities of 74\% and 31\% that the two Seyfert types would differ
by the observed amount if they were drawn
from the same parent population.
The contradiction with Schmitt et al. (2001a), who found
the Seyfert 1 radio sources to be smaller than the Seyfert 2
sources in the 60-$\mu$m sample, may be
due to our larger beam size (by an areal factor of 25 for full-resolution
maps, and more than 100 for tapered maps).  This provides considerably
higher sensitivity to low-surface-brightness emission that is not
necessarily jet-related, and might not 
have been detected in the high-resolution 3.6~cm study of 
the 60-$\mu$m sample.  There also is some possibility (see \S~4.3)
that the Seyfert~1 galaxies in the Palomar sample are selected from
higher on the luminosity function.  This could cause a bias toward
larger Seyfert~1 radio sources, if radio sizes are correlated with
radio power, as reported in some previous work (Ulvestad \& Wilson 1984b).

\vskip 0.3cm

\psfig{file=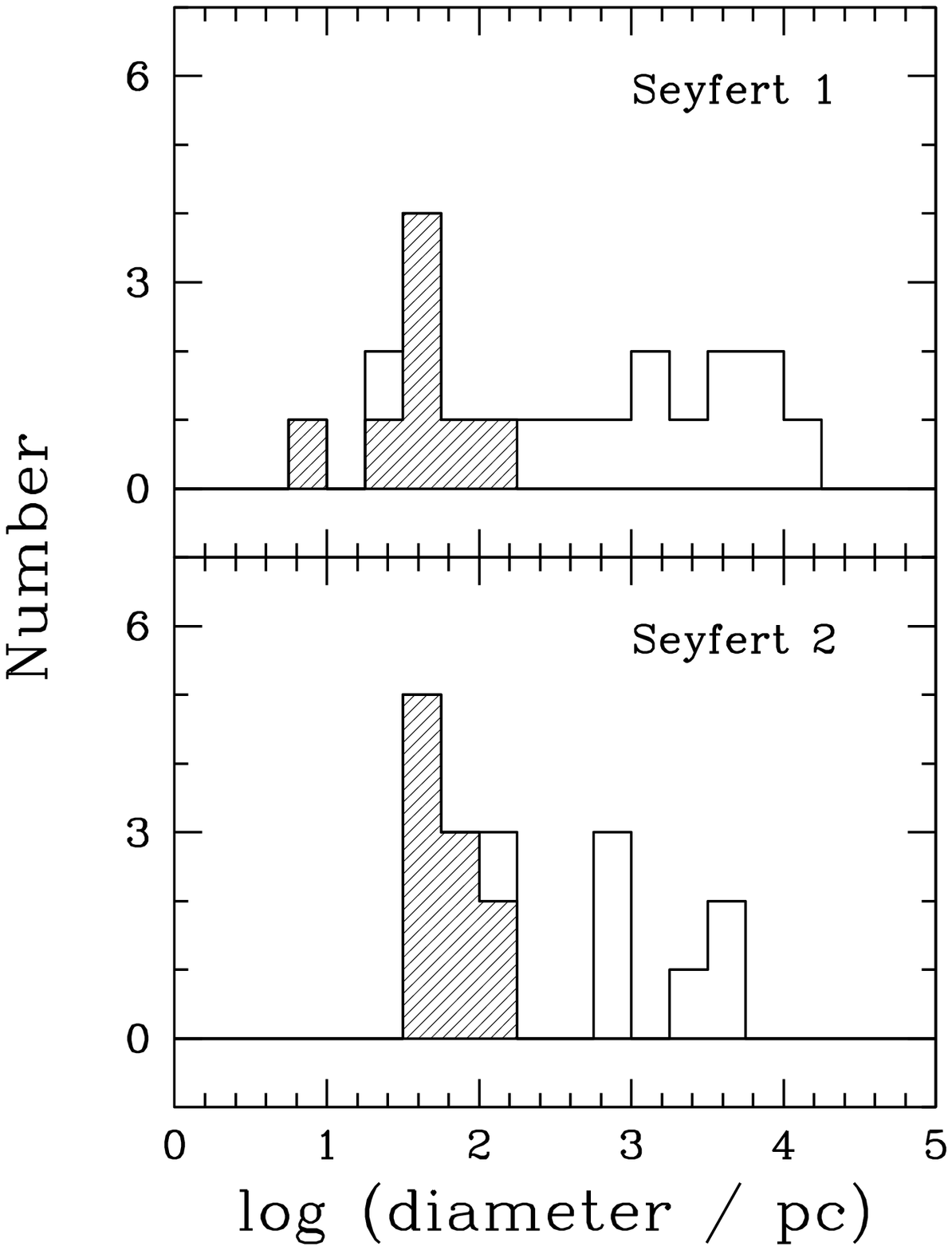,width=8.5cm,angle=0}
\figcaption[f9.eps]{
Histogram of radio sizes for the Palomar Seyfert
sample, for the emission inferred to be associated with
the active nucleus.  Upper limits are shown hatched.}
\label{fig:rsizes}

\vskip 0.3cm

\subsection{Radio Morphology}
\label{sec:morph}

The radio morphologies of the Palomar Seyferts were classified by
Ho \& Ulvestad (2001) according to the scheme introduced by 
Ulvestad \& Wilson (1984a).
Eleven of the 37 detected sources from the statistical sample of 
45 galaxies contain 
linear radio sources (class ``L''), while 18 are only 
unresolved or slightly resolved (classes ``U'' and ``S,''
respectively).  An additional six galaxies were classified as 
having ambiguous (class ``A'') structures by Ho \& Ulvestad (2001), but all
of these sources are barely detected, and the detected
components are poorly resolved.  The fraction of linear
radio sources (11 of 45, or 24\%) can be compared with
other samples of comparable size.  In the heterogeneous 
distance-limited sample of Ulvestad \& Wilson (1984b, 1989), 13 of 57
galaxies (55 detected) have linear radio sources, while 20 of 107 
galaxies in the 12-$\mu$m sample (Thean et al. 2001) have linear sources.
This indicates that about 20\% to 30\% of Seyfert galaxies contain
linear radio sources detectable by current connected-element
interferometers.  However, measurements of radio morphologies
and sizes may be biased by the fact that the dynamic range
on the weakest galaxies is too low to see the full extent of the
radio emission (Ulvestad \& Wilson 1989; Thean et al. 2001).  
Indeed, Thean et al. (2001) find
that the fraction of linear sources in the 12-$\mu$m sample 
rises from 18\% to 28\% when only the galaxies with 3.6~cm flux
densities above 5~mJy are considered.  

Another way of assessing the true fraction of linear sources
is to see whether the apparent radio morphologies
are correlated with radio power.  If we take the 11 galaxies in the
Palomar statistical sample that are classified as linear {\it vs.} the 18
objects that are either slightly resolved or unresolved, 
a Gehan-Wilcoxon test shows only that the two
sets of objects differ in 6~cm radio luminosity at the 8.8\% significance
level.  If we then add the six sources classified 
as ambiguous by Ho \& Ulvestad (2001) to the poorly resolved class
(more consistent with previous work), distributions drawn from the same
parent population would differ by the observed amount only 1.4\%
of the time.  This result
indicates that discussions of Seyfert radio morphologies should
be carried out with caution, since it may well be that the
weaker radio sources also are linear, but are so weak that the
extended emission has not been detected.  This possibility could
be tested with observations that are 10 times more sensitive,
such as those that could be made with the proposed Expanded
Very Large Array (Perley 2000).

A final test of interest is to compare the radio morphologies
to the Seyfert types.  According to the unified scheme, where the 
Seyfert 1 radio sources would be seen more end-on, 
more linear radio sources might be distinguishable
in Seyfert 2 galaxies.  In \S~3.7, it was shown
that there was no significant correlation between radio size and
Seyfert type for the Palomar sample.  The Gehan-Wilcoxon
test shows that for the galaxies classified as linear {\it vs.} 
the slightly resolved or unresolved galaxies, the
morphological classes for the two Seyfert types would differ by the
observed amount 9.2\% of the time if they were drawn from the 
same parent sample.  However, 
this probability shrinks to 2.9\% if the six ambiguous objects
are included with the poorly resolved sources.  
The difference in the Palomar sample is in the sense that the
Seyfert 1 galaxies have a {\it higher} probability of containing
linear radio sources than the Seyfert 2 galaxies, opposite to
what might be expected if the Seyfert 1 radio sources are seen
more end-on.  However, for the small number of linear sources in
the sample, and given a possible tendency to identify stronger
Seyfert 1 galaxies (see \S~4.3), this result 
requires verification in a larger sample.

\subsection{Radio Axes vs. Galaxy Axes}
\label{sec:axes}

Seyfert galaxy radio sources are thought to be powered by radio
jets which emerge from the vicinity of a supermassive black hole,
in a direction perpendicular to a circumnuclear torus.  If the gas in that 
torus is accreted from the large-scale disk of the host galaxy, it should 
share the host's angular momentum axis.  Therefore, one might expect the 
radio jets to appear predominantly perpendicular to the plane
of the host galaxy.  An early study (Ulvestad \& Wilson 1984b) indicated
that there was no excess of Seyferts with jets perpendicular to
the host disks; this study was extended and confirmed by
later work on larger samples, which also took account of the
effects of projection of jets and galaxies on the 
sky (Schmitt et al. 1997; Nagar \& Wilson 1999; Kinney et al. 2000).

In the Palomar Seyfert sample, there are 15 galaxies having
measurable radio major axes (Ho \& Ulvestad 2001).  Figure~10 shows
a histogram of the position angle difference between the
radio and host galaxy major axes of these 15 galaxies.  
Clearly, the distribution is consistent with being uniform.  As discussed by
(Kinney et al. (2000), this is expected in a 
situation in which the radio jets are oriented completely
randomly with respect to the host galaxies.  If the radio jets
are perpendicular to the central accretion disk or torus, this
result implies that the disk/torus does not share the angular
momentum axis of the host galaxy, perhaps implying an external
origin for the accreting gas.  More detailed discussion of
possible reasons for the observed distribution of position angle
differences can be found in Kinney et al. (2000).

\vskip 0.3cm

\psfig{file=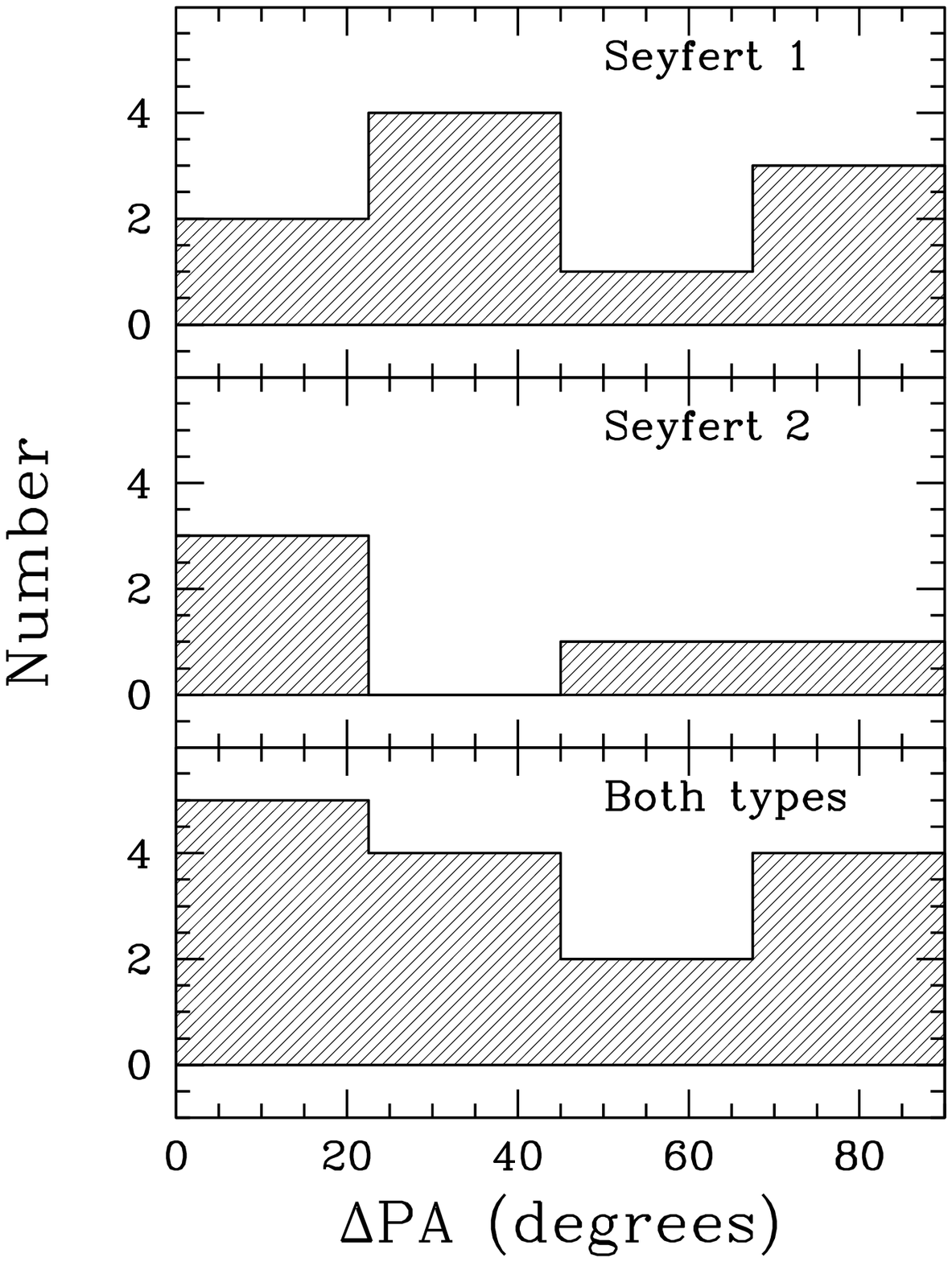,width=8.5cm,angle=0}
\figcaption[f10.eps]{
Histogram of the difference between the position angle of the 
radio and host galaxy major axes for 15 well resolved objects in the 
Palomar Seyfert sample.  The radio axis is from the emission inferred
to be associated with the active nucleus, rather than with
the large-scale galaxy.}
\label{fig:axes}
\vskip 0.3cm

\section{Additional Interpretation}
\label{sec:interp}

A considerable amount of discussion about the various statistical 
correlations in Seyferts has been carried out in \S~3.  The 
physical meaning of various correlations has been explored at
length recently by Thean et al. (2001) for the 12-$\mu$m Seyfert sample,
and by (Schmitt et al. 2001a) for the 60-$\mu$m Seyfert sample;
we do not propose to conduct a similar exhaustive discussion for the
Palomar Seyferts.  Instead, we limit ourselves to some general
points, and a more detailed exploration of some of the results that
appear unique to the Palomar sample.

First, we summarize the results of our statistical tests in 
Table~3.  For each test described in \S~3, the statistical 
significance of that test is given, along with a reference to the relevant
subsection.  For completeness, the table also includes a few additional tests 
that are discussed below.  Customarily, the results of statistical 
tests are deemed significant if the null hypothesis can be rejected at 
a probability of 5\% or less.  In Table~4, we list the primary 
arcsecond-resolution surveys of various Seyfert samples.
Statistical results for these samples are of 
interest for comparison with the Palomar Seyferts, and will occasionally
be cited below.

\subsection{The Space Density of Weak Seyfert Galaxies}
\label{sec:density}

The optical luminosity function given in Table~1 can be 
integrated to find the total space density of weak Seyfert galaxies, such as 
those which dominate the Palomar sample.  In the range of 
$-22$ mag $\leq M_{B_T} \leq -18$ mag, the space density of Palomar 
Seyferts is $(1.25\pm 0.38)\times 10^{-3}$~Mpc$^{-3}$.  By comparison, we can 
look at the local space density of bright quasars, as derived by
Hewett, Foltz, \& Chaffee (1995).  
In the range $0.2 < z < 0.5$ (after conversion to our
Hubble constant), they find space densities for bright quasars 
of $\sim 5\times 10^{-6}$~Mpc$^{-3}$, or $\sim 1.4\times 10^{-6}$~Mpc$^{-3}$
after the objects with Seyfert galaxy luminosities are eliminated.
Thus, the local space density of weak Seyferts appears to be
$\sim 700$ times higher than the local space density of bright quasars.

As a consistency check, we also can compare the Seyfert results to 
the overall space density of galaxies in the Palomar galaxy survey.
First, we eliminate the galaxies having southern declinations or
apparent magnitudes $B_T > 12.64$~mag, which reduces the initial
sample of 486 galaxies (Ho et al. 1995) to a statistical sample of
435 galaxies.  Computation of the space density of bright galaxies
then follows the $V/V_{\rm max}$ method.  We find that the total
space density of bright galaxies is $(3.0\pm 1.2) \times 10^{-1}$~Mpc$^{-3}$.
This density is dominated by intrinsically faint, nearby galaxies.
Using only the 392 galaxies with absolute magnitudes
of $-22\ {\rm mag} \leq M_{B_T} \leq -18\ {\rm mag}$, where the density 
of Seyfert galaxies was derived, we find a cumulative space density of
$(1.30\pm 0.11) \times 10^{-2}$~Mpc$^{-3}$ for the Palomar bright
galaxies.  This is in good agreement with 
the value found for the RSA spirals by Condon (1989), after
conversion to our distance scale.  Thus, Seyferts comprise 9.6\%$\pm$3.0\% of
the space density of nearby galaxies brighter than $M_{B_T}=-18$ mag, as
expected from the fact that 45 of the 392 objects (11.5\%) above 
this optical luminosity are members of our statistical Seyfert sample.
We also can compare the space density of Palomar Seyferts to that of the CfA
Seyferts (Huchra \& Burg 1992), after conversion to our distance scale.
The derived CfA Seyfert space density in the range 
$-22$ mag $\leq M_{B_T} \leq -18$ mag 
is $(1.47\pm 0.39)\times 10^{-4}$~Mpc$^{-3}$.  Formally, the ratio of the
CfA Seyfert density to the Palomar Seyfert density is then
$0.12\pm 0.05$.  The overall galaxy density in the CfA sample is 
$\sim 1.1\times 10^{-2}$~Mpc$^{-3}$ 
in the same absolute magnitude range (comparable to the RSA value
found above), so the fraction of galaxies detected as Seyferts
in the CfA survey is only $\sim 1$\%.  This indicates, as discussed
by Ho \& Ulvestad (2001), that the CfA survey identified only a small fraction
of the galaxies containing weak Seyfert characteristics in their 
optical spectra.

\subsection{Emission-Line Correlations}
\label{sec:emission}

For all the Seyfert samples observed at arcsecond resolution, and 
tested statistically, the radio source powers are strongly correlated
with the strengths and widths of the narrow optical emission lines. Whittle 
(1992a, 1992b) has made a detailed study of the origin of
these correlations.  He found that most Seyfert galaxies have 
emission-line widths that are virialized in the bulge potential of the host 
galaxy.  However, the Seyferts with more powerful linear radio sources have 
``extra'' line width, which can be attributed to interaction between the 
line-emitting gas and the moderately strong radio
jets in these objects.  The latter interpretation is supported
by comparison of Seyferts with linear radio morphologies
to high-resolution emission-line imaging of the same objects; the
radio sources and emission lines have similar angular extents and are roughly 
aligned along the same position angles (Whittle et al. 1986; Wilson \&
Tsvetanov 1994; Bower et al. 1995; Capetti, Axon, \& Macchetto 1997;
Falcke, Wilson, \& Simpson 1998).  This 
spatial relationship could be due to interactions between the emission-line 
clouds and the radio plasma, although one also might expect such a result 
based on collimation of the radio jets and generation of a 
photoionization cone by the same accretion disk or torus.
A detailed study by Bicknell et al. (1998) indicates that the narrow-line region
might be completely powered by the radio jet, although this study uses 
as examples only several relatively high-power linear radio sources, and 
not the bulk of the less radio-luminous Seyferts.  The overall narrow 
emission-line spectrum of Seyferts can be reasonably explained in the 
context of pure photoionization models (e.g., Ferland \& Osterbrock 
1986; Ho, Shields, \& Filippenko 1993b).  In addition, reverberation 
mapping of the smaller broad-line region (Peterson 1993; Wandel, 
Peterson, \& Malkan 1999; Peterson \& Wandel 2000)
strongly favors the standard photoionization paradigm with an ionizing
continuum similar to that invoked to account for the narrow lines.

For a given optical line strength or width, the radio power of the
Palomar Seyferts may range over two orders of magnitude or more,
as shown in \S~3.5.  This suggests that the emission
lines and radio-emitting plasma may not be directly related, but
are instead controlled by some third governing factor, such as the
properties of the nuclear bulge (Whittle 1992a, 1992b).
Recent work by Evans et al. (1999) indicates that the ionization of the
gas does not seem to be strongly influenced by pre-ionization from
the radio jet in Seyfert galaxies, so the jet appears to be coupled
only loosely to the line-emitting gas.  

In the Palomar Seyfert sample, we find that the {\it ratio}\ of radio
power to \oiii\ luminosity is significantly correlated with 
Seyfert type: it is higher in Seyfert~1 galaxies than in 
Seyfert~2 galaxies (see \S~3.5).  Careful inspection
of Figure~5 shows that this difference may be 
strongly influenced by the low-luminosity objects; the Seyfert 1 and
Seyfert 2 galaxies appear to be well mixed within those objects
having \oiii\ luminosities above $L$(\oiii) = $10^{39.5}$~erg~s$^{-1}$,
while the Seyfert 2 galaxies seem to have systematically
weaker radio sources for lower $L$(\oiii).  Accordingly, we have divided 
the galaxies into two samples, those with $L$(\oiii) above and below 
$10^{39.5}$~erg~s$^{-1}$.  Application of the Gehan-Wilcoxon test to these two 
samples shows that there is no significant correlation of the radio/\oiii\ ratio
with Seyfert type among the galaxies with the high values of $L$(\oiii);
probabilities that the two Seyfert types would differ by as much as
observed if they were drawn from the
same parent sample are 51\% for the 20~cm power and 26\% for the
6~cm power.  However, for the weaker objects, the similar probabilities 
are only 1.3\% at 20~cm and 3.1\% at 6~cm.  

The differences between Seyfert 1 and Seyfert 2 galaxies among
the objects with low emission-line luminosities
would be even more significant were it not for two galaxies with 
high radio powers and only upper limits to their \oiii\ emission,
shown most clearly in the upper left of Figure~5{\it b}.
It turns out that these two objects are NGC~777 and NGC~4472,
two of the four elliptical galaxies in our statistical sample.  It is 
conceivable that the radio-\oiii\ relation depends on the circumnuclear 
environment of the host, which might differ between elliptical and spiral
galaxies.  If we remove all four elliptical galaxies from the sample, the
probabilities that the low-luminosity Seyfert 1 and Seyfert 2 galaxies
would differ by the observed amount if 
drawn from the same parent population are only 0.1\% at 20~cm
and 0.6\% at 6~cm.  The median value of 
$\log~[P_{20}$~(W~Hz$^{-1})/L$(\oiii)~(erg~s$^{-1})]$ is $-19.1$ for 
Seyfert 1 galaxies and $-20.3$ for Seyfert 2 galaxies; these were calculated 
using the Kaplan-Meier product-limit estimator (Feigelson \& Nelson 1985) 
to take into account the upper limits.

A test for the possible influence of starbursts is to see 
whether the weak Seyfert~2 galaxies follow the rather tight 
empirical correlation
between 20~cm radio flux density and {\it IRAS}\ far-infrared flux 
(Helou, Soifer, \& Rowan-Robinson 1985; Condon, Anderson, \&
Helou 1991; Condon 1992).  The parameter $q$, which describes 
this ratio, is defined (Helou et al. 1985) as 
\begin{equation}
q \equiv \log\biggl({FIR\over 3.75\times 10^{-12}\ {\rm W\ m}^{-2}}
\biggr)\ -\ \log\biggl({S_{20}\over {\rm Jy}}\biggr)\ ,
\end{equation}
where
\begin{equation}
FIR\ \equiv\ 1.26\times 10^{-14}\ 
\biggl({2.58S_{60\mu{\rm m}}\ +\ S_{100\mu{\rm m}}\over {\rm Jy}}
\biggr) {\rm W\ m}^{-2}; 
\end{equation}

\noindent $S_{20}$, $S_{60\mu{\rm m}}$, and $S_{100\mu{\rm m}}$ are the
flux densities at 20~cm, 60~$\mu$m, and 100~$\mu$m, respectively.  For starburst 
galaxies, $q$ typically takes on the values $2.3\pm 0.2$.  

In the Palomar Seyfert statistical sample, 43 of the 45 galaxies have 
measured \oiii\ luminosities and values or limits to $q$, based on {\it IRAS}\ 
flux densities given in Ho et al. (1997a); these quantities are plotted 
against one another in 
Figure~11.  There is an obvious inverse correlation between
$q$ and $L$(\oiii), which is expected since the definition of $q$ involves 
radio luminosity, known to be strongly correlated with $L$(\oiii) 
(Fig.~5); 
computation of the generalized Kendall's $\tau$ coefficient shows that
this correlation is significant at the 0.01\% level.   In addition, 
Kendall's $\tau$ shows that $q$ is correlated with Seyfert type, with
only a 1.3\% probability that the Seyfert 1 and 
Seyfert 2 galaxies are drawn from the same parent population.  An important 
point to note from Figure~11 is that the Seyfert 2 galaxies 
with less luminous emission lines [$L$(\oiii) $<$ $10^{39.5}$~erg~s$^{-1}$]
almost all have $q\geq 1.8$ and have no detections at 20~cm.
Thus, the infrared strengths of the low-luminosity Seyfert 2 galaxies
are consistent with the possibility that much of their infrared
and (undetected) 20~cm radio emission are due to starbursts.  This
is not a unique conclusion; for example, it also appears that radio-quiet
quasars follow a similar radio-infrared relation to starbursts, but 
that the similarity could be entirely coincidental (Sanders et al. 1989).

\vskip 0.3cm

\psfig{file=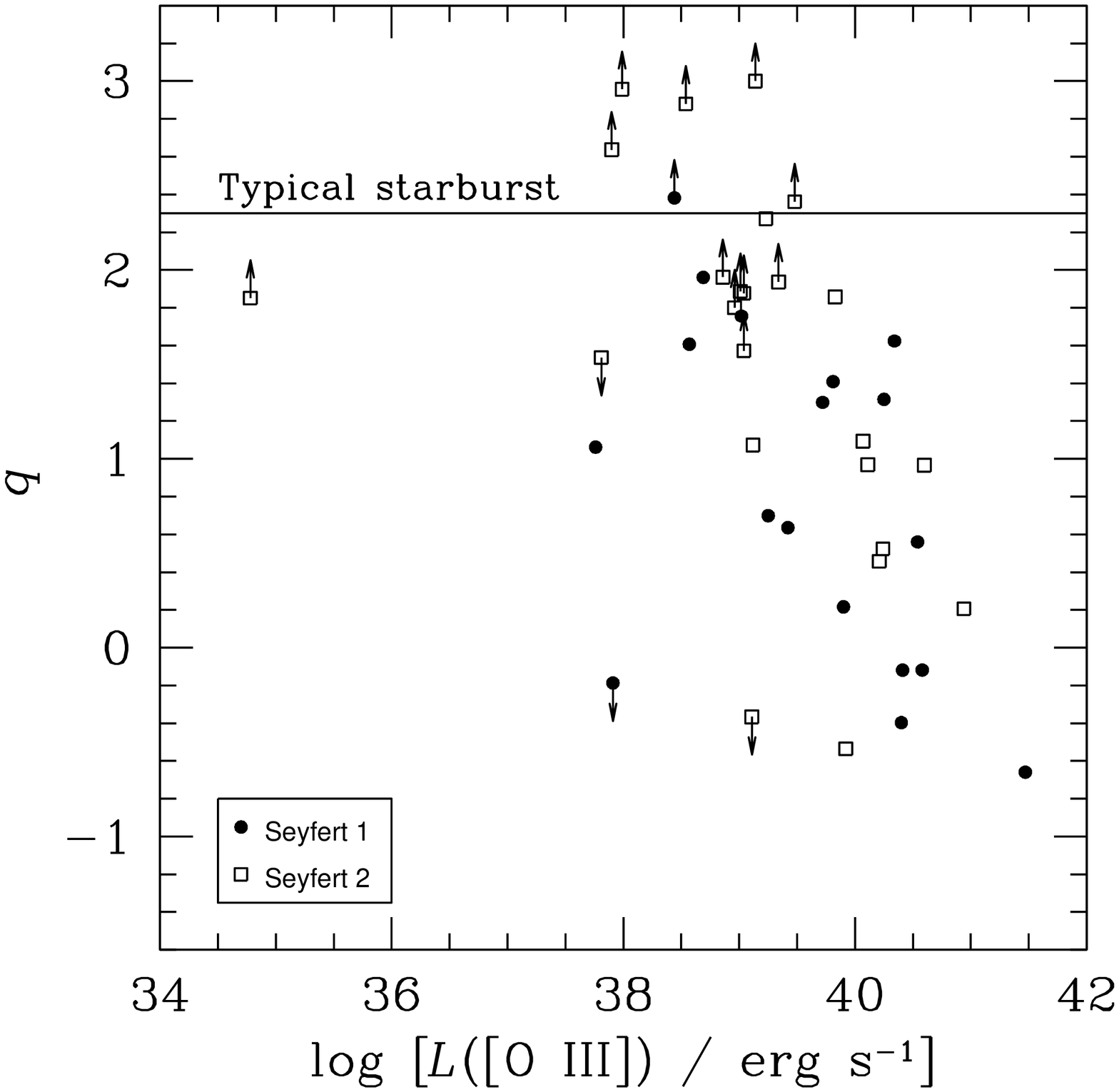,width=8.5cm,angle=0}
\figcaption[f11.eps]{
{Plot of $q$, the logarithmic ratio of far-infrared
flux to 20~cm flux density (defined in text), {\it vs.} the extinction-corrected
\oiii\ luminosity, for the Palomar Seyfert sample.}
\label{fig:qoiii}}
\vskip 0.3cm

Another way to consider whether star formation can account for the
properties of the weaker Seyfert~2 galaxies is to compare in greater 
detail their relative emission-line strengths.  
If, for instance, the low-$L$(\oiii) objects
are systematically contaminated by emission from \hii\ regions, we 
would expect them to exhibit, on average, weaker low-ionization forbidden 
lines of \oi\ $\lambda\lambda$6300, 6364, \nii\ $\lambda\lambda$6548, 6583, 
and \sii\ $\lambda\lambda$6716, 6731, compared to high-$L$(\oiii) objects, 
since O-star photoionization generates characteristically weak
low-ionization lines (Veilleux \& Osterbrock 1987; Ho, Filippenko,
\& Sargent 1993a).  However, the 
diagnostic diagrams plotted in Figure~12 reveal no clear 
differences between the two subsets of objects.


\begin{figure*}[t]
\vspace{-2.7cm}
\centerline{\psfig{file=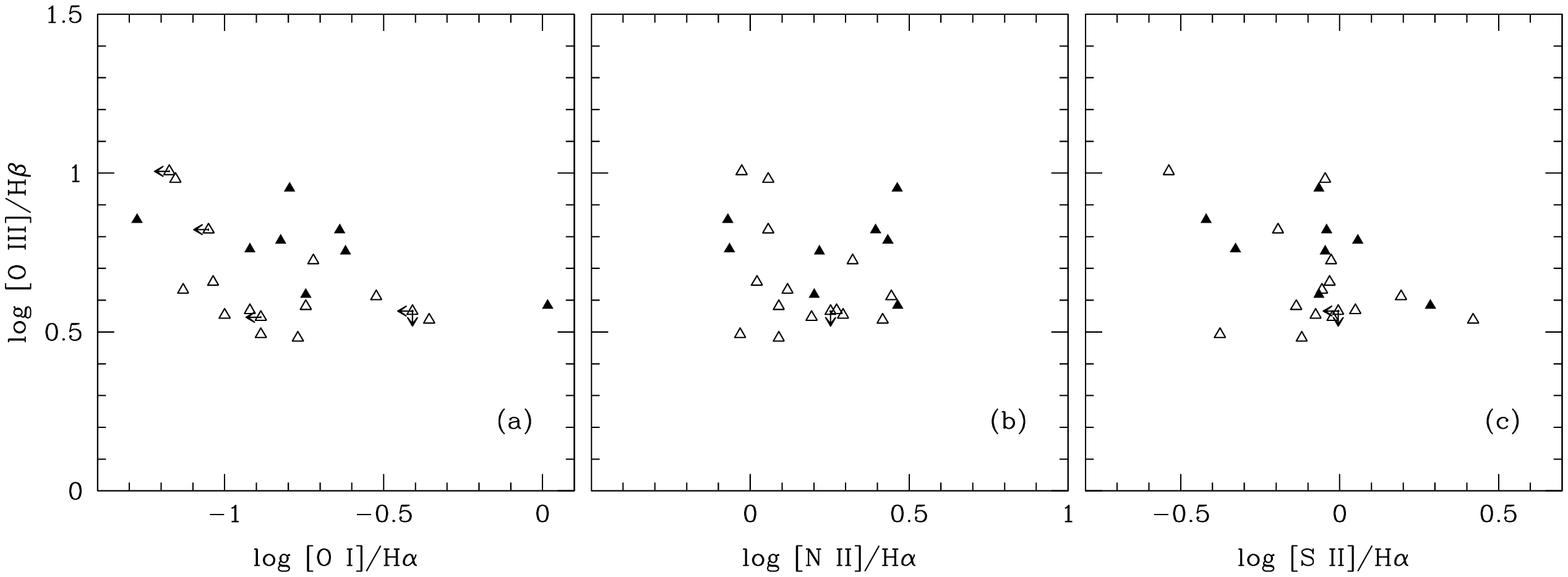,width=18.5cm,angle=270}}
\vspace{-3.5cm}
\figcaption[f12.ps]{
Diagnostic line-ratio diagrams for the 25 Seyfert 2 galaxies in 
the Palomar sample.  The three panels plot the logarithmic ratios of 
\protect\oiii\ $\lambda$5007/H$\beta$ {\it vs.}  ({\it a\/}) 
\protect\oi\ $\lambda$6300/H$\alpha$, 
({\it b\/}) \protect\nii\ $\lambda$6584/H$\alpha$, and 
({\it c\/}) \protect\sii\ $\lambda\lambda$6716, 6731/H$\alpha$.  
Objects with $L$(\protect\oiii) $\geq$ $10^{39.5}$~erg~s$^{-1}$ 
and $L$(\protect\oiii) $<$ $10^{39.5}$~erg~s$^{-1}$ 
are shown with filled and open symbols, respectively.  There are 
no systematic differences between the two subsets of objects.
\label{fig:oiiinii}}
\end{figure*}
\vskip 0.3cm

There are several possible explanations for the anomalous 
weakness of the radio emission from the low-luminosity Seyfert~2
galaxies.  First, the galaxies may be dominated by a starburst component.  
The absence of spectral differences in the optical (Fig.~12),
and the low radio-\oiii\ ratio 
suggest that this is unlikely.  Second, Seyfert galaxies may require a 
certain minimum level of nuclear activity to ``turn on'' their radio
emission.  According to the unified scheme, this would mean that 
Seyfert~1 galaxies with low forbidden-line luminosities would be expected to
have radio emission as weak as the Seyfert~2 objects.  However,
at the low forbidden-line strengths, the Seyfert~2 galaxies seem to
have weaker radio sources than the Seyfert~1 galaxies.
Third, the weak Seyfert~2 galaxies may have their nuclear radio
emission obscured in some way, a possibility that is
explored further in \S~4.4.

\subsection{Radio Correlations with Seyfert Type}
\label{sec:radtype}

For the Markarian Seyferts (de Bruyn \& Wilson 1978; Ulvestad \&
Wilson 1984a) and the first
distance-limited sample of Seyferts (Ulvestad \& Wilson 1984b), it appeared
that type 2 Seyferts were stronger radio sources than type 1 
Seyferts.  However, samples that were selected more uniformly
and imaged at arcsecond resolution 
have never shown a convincing difference in the radio powers between the
two types (Ulvestad \& Wilson 1989; Thean et al. 2001; Schmitt et al. 2001a), 
nor have most heterogeneous
compilations of Seyferts (Giuricin et al. 1990).  Thus, it came as a
distinct surprise to find a statistically significant result 
 that Seyfert 1 galaxies appear to have somewhat stronger
radio sources than Seyfert 2 galaxies (see Table~3
and \S~3.4.1).  As discussed above (\S~4.2), this result seems
to be caused by anomalously weak radio emission (for a given \oiii\ 
luminosity) in the low-luminosity Seyfert~2 galaxies.

Given the manner in which the optical classifications are made, 
we must consider whether subtle selection effects are inherently built
into the Palomar Seyfert sample, or indeed into any sample selected by 
optical spectroscopy.  As shown in \S~3.1, there is no
significant difference in the distance between the Seyfert types.  
However, the detection of broad emission lines, on which the distinction 
between type~1 and type~2 Seyferts is based, depends critically on the 
strength of the line emission relative to the continuum (the equivalent 
width), for spectra of fixed signal-to-noise ratio.  This is especially 
true for Seyfert galaxies with intermediate types 1.8 or 1.9, which 
dominate the type 1 objects identified in 
the  Palomar sample, because of the challenge in detecting the 
faint, low-contrast wings of the broad component of the H$\alpha$ line  
(Ho et al. 1997c).  All else being equal, we expect broad H$\alpha$ to be 
more easily recognized in objects with higher line luminosities and 
higher line equivalent widths.  Figure~13{\it a} indicates
that the weak and strong Seyfert~2 galaxies show no distinction in
their level of excitation, as characterized by the \oiii/H$\beta$ ratio.
However, as shown in Figure~13{\it b},
the Seyfert~1s in the Palomar sample do indeed have higher 
H$\alpha$ equivalent widths than the Seyfert~2s.  The difference in line
luminosity is only marginal for \oiii\ (Table~3), but it is more significant 
for H$\alpha$ (probability of 0.8\% that the two Seyfert types are drawn from
the same parent population).  
The median extinction-corrected luminosity of the 
narrow-component of H$\alpha$ is 5.0$\times 10^{39}$~erg~s$^{-1}$ for the 
Seyfert~1s, to be compared with 1.4$\times 10^{39}$~erg~s$^{-1}$ for the 
Seyfert~2s.  Since radio power scales with line luminosity 
(Fig.~5; see discussion in Ho \& Peng 2001), we naturally 
expect the Palomar Seyfert~1s to be on average more radio luminous than the 
Seyfert~2s.


\begin{figure*}[t]
\vspace{-2.2cm}
\centerline{\psfig{file=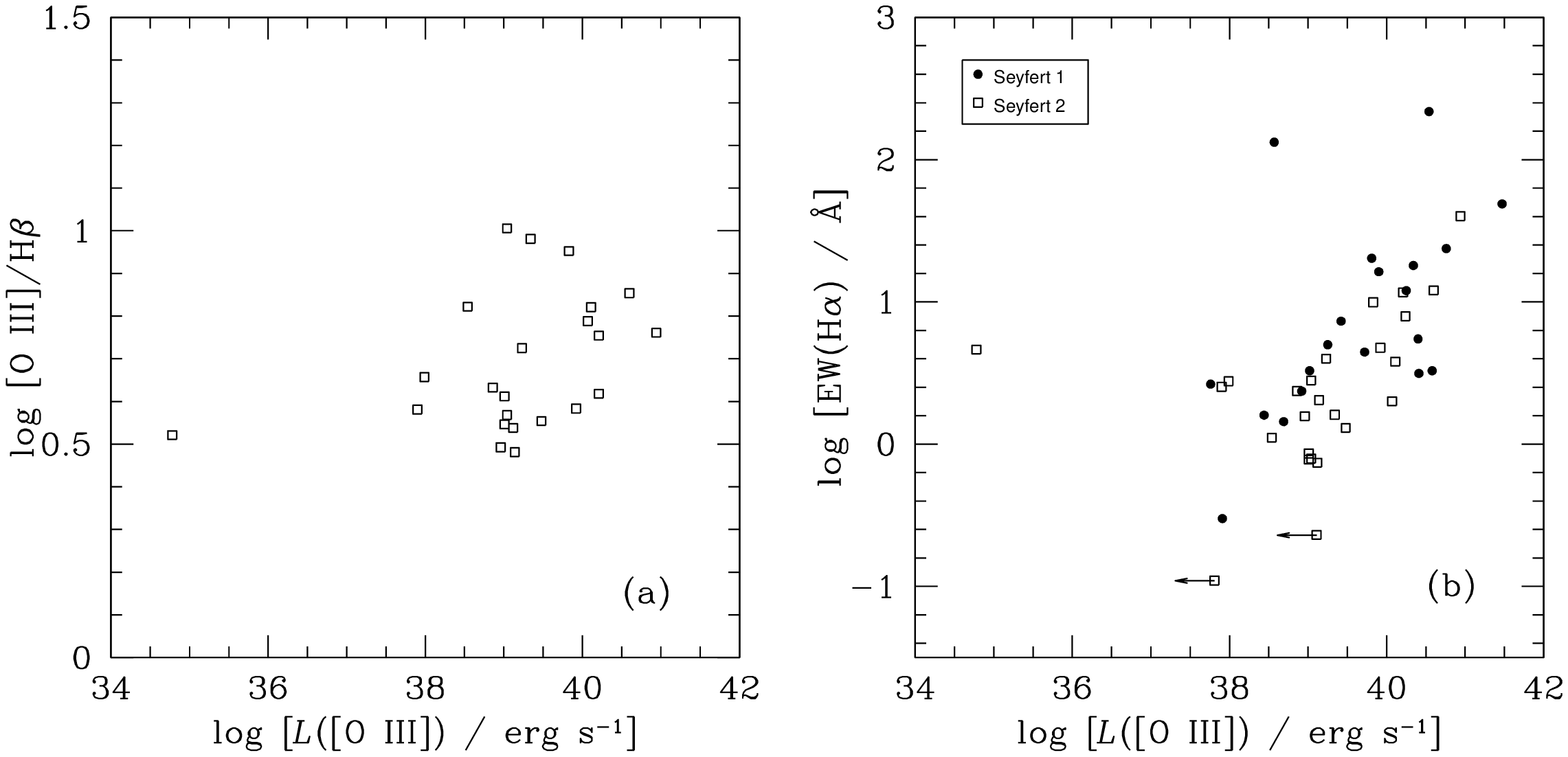,width=18.5cm,angle=270}}
\vspace{-3.3cm}
\figcaption[f13.ps]{
Emission-line properties plotted against the 
extinction-corrected \protect\oiii\ luminosity for Seyfert galaxies 
in the Palomar sample.  ({\it a\/}) \protect\oiii/H$\beta$ ratio is plotted 
for the Seyfert 2 galaxies (two elliptical galaxies with 
\protect\oiii\ upper limits are excluded).  ({\it b\/}) Equivalent width 
of the narrow component of the H$\alpha$ emission 
line is plotted for all the Palomar Seyfert galaxies.
\label{fig:oiii}}
\end{figure*}
\vskip 0.3cm

\subsection{The Flat-Spectrum Seyferts}
\label{sec:flat}

In \S~3.6, it was shown that a substantial 
fraction of the Palomar Seyferts have relatively flat radio spectra. 
This fraction appears considerably higher than in the
distance-limited sample of Ulvestad \& Wilson (1984b, 1989), even
after accounting for possible resolution effects.  
The other samples listed in Table~4 do not have 
radio spectra determined at arcsecond resolution, so they cannot
be usefully compared.  There are several possible explanations for
the flat-spectrum objects.  First, the
Palomar Seyferts may have objects that contain a significant
fraction of their emission from supernova remnants like those
in our Galaxy.  Second, they may be synchrotron emitters
that include self-absorbed radio cores, causing the overall spectra
to be flatter than optically thin sources.  Third, the emission
may be ``contaminated'' by thermal gas, which either free-free absorbs
steep-spectrum synchrotron emission, or contributes its own
intrinsic emission with a nearly flat spectrum.  
We consider each of these possibilities in turn in
the following subsections.

\subsubsection{Supernova Remnants}

The Palomar Seyfert radio emission conceivably could arise from a 
collection of supernova remnants rather than from an active
galactic nucleus.  However, there are several strong arguments
against this possibility.  Collections of extragalactic supernova 
remnants in starbursts appear to have median spectral indices in the 
range between $-0.5$ and $-0.7$ (Lacey, Duric, \& Goss 1997;
Ulvestad \& Antonucci 1997; Hyman et al. 2000; Neff \& Ulvestad 2000), 
while
Galactic supernova remnants have median values of 
$\alpha\approx -0.45$ (Clark \& Caswell 1976).  A substantial 
number of the Palomar Seyferts have much flatter spectra, 
with $\alpha \geq -0.2$ (see \S~3.6).  Therefore,
supernova remnants do not seem able to explain the large fraction
of flat spectra; if they exist in the weak Palomar Seyferts, their
relatively steep spectra must be balanced by a comparable amount
of flat or inverted-spectrum emission from an AGN or a collection
of thermal radio sources.

In addition, as shown in \S~4.2, most Palomar Seyferts do not 
follow the nearly universal correlation between far-infrared and radio 
emission that is expected for galaxies undergoing active star formation.  
Finally, the optical spectra of these sources, by definition, do not 
resemble the spectra of \hii\ regions or of supernova remnants (see, e.g., 
Ho et al. 1993a).  Hence, the possibility that radio emission from many
Palomar Seyferts is affected significantly by supernova remnants 
is not considered further here.

\subsubsection{Self-Absorbed Synchrotron Emission}

If a substantial fraction of the Palomar Seyferts contain 
self-absorbed synchrotron sources (Williams 1963), they must 
have radio cores with brightness temperatures of 
$\sim 6\times 10^9$~K or greater.
A number of the objects with relatively flat spectra are detected
at 6~cm with flux densities of $\sim 0.5$~mJy, and have no detection
at 20~cm.  For an object to have a flux density of 0.5~mJy and
a brightness temperature above $6\times 10^9$~K at 5~GHz, its angular 
size must be less than about 2~mas, corresponding to a
linear size of less than 0.2~pc at a typical distance of 20~Mpc.
This combination of parameters is at the limit accessible by
present-day VLBI systems.  However, 
there are a few stronger flat-spectrum sources in our sample.
A source such as NGC~4565, with a flux density of 2.7~mJy, 
would have a brightness temperature above the self-absorption
limit if its size were less than about 4~mas. 
Such a source could be imaged using the Very Long Baseline Array (VLBA).

We note here that Falcke et al. (2000) have used VLBA 
observations to find a substantial population of high-brightness 
objects among LINERs with cores of a few mJy
or higher, which would favor the possibility that the Palomar Seyferts
could have similar cores.  Therefore, at least one class of weak AGNs has
cores compact enough that they might be self-absorbed, lending some
credence to the possibility that weak Seyferts also might have such
cores.  Such compact sources also would be expected to be variable.  
The only search for variability in a significant sample of weak
AGNs was reported recently by Falcke et al. (2001), and at least some
mJy-level Seyferts from the Palomar sample appear to vary
at 15~GHz on time scales of years.  Radio-quiet quasars with 
flat-spectrum cores, at similar flux levels (but higher
luminosities) appear to have a high incidence of variability as
well (Barvainis, Lonsdale, \& Antonucci 1996).  Further studies of 
radio variability in the Palomar sample would be useful.

The possibility of self-absorption, however, does not explain why the
Palomar Seyferts have significantly flatter spectra than the 
distance-limited sample of Ulvestad \& Wilson (1984b, 1989), in the absence of a 
plausible physical mechanism by which less luminous radio sources, which 
typify the Palomar sample, are systematically more likely to be self-absorbed.
One possibility is that low-luminosity AGNs experience highly 
sub-Eddington accretion rates, such that their accretion undergoes an
optically thin advective flow.  Advection-dominated accretion flows (ADAFs; 
see Narayan, Mahadevan, \& Quataert 1998 for a review) have been invoked to 
explain a number of low-luminosity galactic nuclei (e.g., Narayan, Yi, \& 
Mahadevan 1995; Lasota et al. 1996; Quataert et al.  1999; Ho et al. 2000).
Wrobel \& Herrnstein (2000) have found that the low-power radio cores of some elliptical 
galaxies are consistent with the presence of ADAFs.
ADAF models robustly predict that the self-absorbed radio spectrum should rise 
with frequency as $\nu^{2/5}$ until it reaches a maximum frequency, beyond 
which the plasma becomes optically thin (e.g., Mahadevan 1997; Narayan 
et al. 1998).  The observed spectral indices, therefore, could easily 
arise from an ADAF core component diluted (in some cases) 
by a steep-spectrum jet component.  ADAF models predict that the spectrum
should continue to rise slowly into the millimeter/sub-millimeter regime,
so sensitive observations at 100~GHz or higher could provide a strong
test of the viability of an ADAF model for the low-luminosity Seyferts.

\subsubsection{Thermal Gas}

Another possibility is that enough thermal gas is present 
in the Palomar Seyferts to account for the flat spectra, either by
free-free emission or by free-free absorption of intrinsically steeper
synchrotron spectra.  Low-frequency flattening in a Seyfert spectrum was first 
detected in NGC~1068 at wavelengths longer than 20~cm by Pedlar et al. (1983).
Most of the spectra in Palomar Seyferts are not strongly 
inverted between 6 and 20~cm.  Therefore, if free-free absorption
is present, the typical galaxy is likely to have
an optical depth of unity at some wavelength between 20 and 6~cm.  
However, there are examples such as NGC~1275 (3C~84; one of the members 
of our sample), whose jet is strongly free-free absorbed on parsec scales 
at wavelengths shorter than 6~cm (Levinson, Laor, \& Vermeulen 1995;
Walker et al. 2000).
The situation may be considerably more complex than that for a single
absorbing component; free-free absorption in
Seyfert galaxies imaged with VLBI sometimes appears to occur in front
of some, but not all, individual radio components (Roy et al. 1998;
Ulvestad et al. 1999).  

The free-free optical depth can be approximated as (Mezger \& Henderson 1967)
\begin{equation}
\tau_{\rm ff}\ \approx\ 8.235\times 10^{-2}T_e^{-1.35}\ {\nu_9}^{-2.1}
{n_e}^2 L_{\rm pc}\ ,
\end{equation}
where $T_e$ is the electron temperature in Kelvins, $\nu_9$ is the observing 
frequency in GHz, $n_e$ is the electron density in cm$^{-3}$, and $L_{\rm pc}$ 
is the path length in parsecs.  Using this equation, and assuming the presence
of ionized gas at $T_e = 10^4$~K, we find that $\tau_{\rm ff} = 1$ at 2--3~GHz 
for an emission measure 
$E\,\equiv\,{n_e}^2 L_{\rm pc}\,=\, (1.3-3.1)\times 10^7$~cm$^{-6}$~pc.

Most of the flat-spectrum radio sources among the Palomar Seyferts
are dominated by an unresolved component, with a typical size upper 
limit of 100~pc or less (see Fig.~9).  As an illustrative example, 
we consider the Seyfert~1.9 galaxy NGC~4565.  This galaxy has a total radio 
flux density of 2.7~mJy at both 6 and 20~cm, as well as an upper limit to its 
diameter of $d$ = 25~pc.  As an initial trial, we assume that the ionized gas is 
confined to a spherical volume of this size, centered on a point radio source.  
Therefore, the maximum path length $L\,=\, d/2\,=$ 12.5~pc.  Assuming uniform 
density along this line of sight, the density required to generate the
estimated emission measure is $n_e$ = 1000--1600~cm$^{-3}$.  However, for an 
ionized region 12.5~pc in radius with this density, the predicted H$\alpha$ 
luminosity for Case~B recombination (Osterbrock 1989) would be at least 
$8.5\times 10^{40}$~erg~s$^{-1}$, far in excess of the observed value of 
$2.8\times 10^{38}$~erg~s$^{-1}$ (Ho et al. 1997a).  Therefore, the simplest
picture is unlikely to be correct.

The above result is a manifestation of the well-known fact that the
narrow-line regions of Seyferts must have filling factors for the ionized
gas that are  much smaller than unity (e.g., Osterbrock 1989, \S 11.5).  In 
order to make the predicted and observed values of its H$\alpha$ luminosity
consistent, it is necessary to decrease $L$ and increase $n_e$
relative to the initial assumption.  Consider
the case where we view the NGC~4565 radio source 
through the center of a single cloud of diameter $L_{\rm pc}$, and 
assume an emission measure of $E=2\times 10^7$~cm$^{-3}$, or a turnover 
frequency of 2.4~GHz.  Then, both the spectral turnover and the
observed H$\alpha$ luminosity could be accounted for by a path length
of 1.6~pc and $n_e=3500$~cm$^{-3}$.  Of course, it
may be that the required amount of thermal gas would be broken up
into a number of smaller clouds which provide a similar total path 
length and volume.  In any event, it appears that free-free
absorption is a possible explanation for the flat spectrum only
if the radio source is considerably smaller than the VLA
upper limit of 25~pc.

As a check for self-consistency, we can predict the total 
free-free emission from
the hypothetical absorbing cloud and compare it to the 
observed power from the NGC~4565 nucleus.  The predicted 
free-free emission over all frequencies, from a 1.6-pc 
cloud with an ionized density of 3500~cm$^{-3}$, would be 
$\sim 1.4\times 10^{38}$~erg~s$^{-1}$.  This luminosity would
be contained in a free-free spectrum that should continue up
to a frequency $\nu\sim kT/h \sim 2\times 10^{14}$~Hz (1.5~$\mu$m
wavelength).  Assuming a thermal spectrum with spectral index
$-0.1$ from the estimated turnover frequency of 2.4~GHz up to
$2\times 10^{14}$~Hz, we predict a free-free flux density from
the cloud of only $\sim 20$~$\mu$Jy at 2.4~GHz, less than 1\% of
the observed value of 2.7~mJy at gigahertz frequencies.
Therefore, the radio data are consistent with the presence of 
the hypothetical obscuring cloud, but that cloud cannot account
for the bulk of the radio emission.  This is similar to the
result found for radio-quiet quasars by Antonucci \& Barvainis (1988). 

The hypothesis of free-free absorption, which appears to be
consistent with the observations to date, also might explain why the
weak Seyfert 2 galaxies appear to be systematically weaker radio
emitters in the Palomar sample than the Seyfert 1 galaxies, 
particularly at 20~cm.  Free-free absorption from either the
nuclear torus or the narrow-line region could be more important
in the low-power Seyfert~2 galaxies.
If absorption is associated with the torus, the apparent excess of
flat- or inverted-spectrum objects in the weak Palomar Seyferts 
might occur if any radio jets present are too weak
to punch their way out into the ionization cone, so that all the
radio emission lies behind or within the torus.  More powerful
Seyferts, such as those in the distance-limited sample of 
Ulvestad \& Wilson (1984b, 1989), would have radio emission extending outside
the torus, and be less likely to be free-free absorbed.  
Using the models of Neufeld, Maloney, \& Conger (1994), 
Wilson et al. (1998) showed that the 
spectral turnover of the Seyfert/LINER galaxy NGC~2639 could be 
accounted for by free-free absorption in the torus, either from a 
warm molecular gas phase or from a hot, weakly ionized atomic gas phase;
the reader is referred to Wilson et al. (1998) for details.
Alternatively, the absorption could be in the narrow-line region.
Weak Seyferts might have more compact narrow-line regions, as appears to
be the case for LINERs (Pogge et al. 2000).  This could provide a larger
covering factor of denser gas for the nuclear radio sources, again
accounting for the putative increased absorption in the weaker objects.

Two possible radio tests exist for the free-free absorption hypothesis. 
First, very sensitive VLA imaging of the weakest radio sources in the 
Palomar Seyfert sample, at 3.6~cm
or shorter wavelengths, could be used to attempt to detect the radio 
emission in the optically thin regime.  In this regard, we note the 
apparent lack of obvious differences between radio properties of 
type 1 and type 2 Seyferts selected from infrared surveys
(Thean et al. 2001; Schmitt et al. 2001a).  Since the radio imaging of 
these samples was done at 3.6~cm, the published images might be
relatively unaffected by free-free absorption, possibly being above 
the frequency of the absorption cutoff for most galaxies.
Second, multi-frequency 
milliarcsecond-resolution VLBA imaging (not possible for the very weakest sources) 
could be used to search for multiple radio components and measure
their spectral shapes.  We recently obtained such VLBA observations of 
three low-luminosity flat-spectrum active galaxies from the Palomar sample
in order to carry out this search.

\section{Are the Data Consistent with the Unified Scheme?}
\label{sec:unified}

We have carried out an analysis of a statistical sample of 45 Seyfert 
galaxies from the Palomar optical spectroscopic survey, for which 
high-resolution radio (VLA) observations were reported by Ho \& Ulvestad (2001).  
The space density of such objects in galaxies having 
$-22$ mag $\leq M_{B_T} \leq -18$ mag 
is $(1.25\pm0.38)\times 10^{-3}$~Mpc$^{-3}$, about 9\%$\pm$3\% of the overall
space density of bright galaxies.  Previous surveys have
identified Seyfert activity in a much smaller fraction of such
galaxies.  A variety of statistical tests have been carried out
on the Palomar Seyfert sample in order to examine the consistency
with the unified scheme for Seyfert galaxies.
Here, we summarize some of the major points that have
been brought out in the discussion and analysis.

First, the following points seem wholly consistent with the unified
scheme:
\begin{enumerate}

\item{The Seyfert 1 and Seyfert 2 galaxies appear in the same types
of host galaxies.}

\item{There is no statistical difference in the forbidden-line
luminosities or widths between Seyfert 1 and Seyfert 2 galaxies.}

\item{For the objects with stronger radio sources, the Seyfert 1 and
Seyfert 2 galaxies obey the same radio-forbidden line relations.}

\item{The radio spectra of Seyfert 1 and Seyfert 2 galaxies are
statistically consistent with each other.}

\item{Seyfert radio axes are uncorrelated with host galaxy axes.}

\end{enumerate}

There are a number of points that, taken at face value, seem inconsistent with
the unified scheme.  However, all have reasonable explanations in the
classical scenario unifying Seyfert 1 and Seyfert 2 galaxies:

\begin{enumerate}

\item{{\it Seyfert 1 galaxies appear to have more luminous radio sources
than Seyfert 2 galaxies.}  There are two possible explanations here.  The
first is a subtle selection bias, in which galaxies can only be identified
as Seyfert 1 galaxies in the Palomar sample, or indeed in any sample selected
by optical spectroscopy, if they have higher emission 
line equivalent widths, implying that the Seyfert 1 objects are identified 
systematically higher on the luminosity function.  The second possible
explanation is that the weak Seyfert 2 galaxies have a strong tendency to
be absorbed.}

\item{{\it Weak Seyfert 2 galaxies have a smaller radio/\oiii\ ratio than
weak Seyfert 1 galaxies.}  This may be explained if a minimum level of
nuclear activity is required to enable a radio jet to emerge from the
vicinity of the central engine, and if the small radio sources below that
minimum activity level are still contained inside or behind the nuclear
torus.}

\item{{\it There is no correlation of radio source size with Seyfert type.}
Only a relatively small fraction (20\%--30\%) of Seyfert galaxies have
resolvable linear radio sources.  The weaker active nuclei prevalent in the
Palomar sample may not generate sources large enough to be resolved by the
VLA, and our relatively small sample does not provide enough linear sources
to expect any statistically significant difference.}

\item{{\it Seyfert 1 galaxies may contain a higher fraction of linear radio 
sources than Seyfert 2 galaxies.}  This is consistent with a possible 
selection effect in which Seyfert 1 galaxies are identified higher
on the luminosity function, if the more powerful objects are more likely
to have linear sources.}

\item{{\it An apparently large fraction of Palomar Seyferts have flat spectra,
when compared with previous samples of more powerful Seyferts.}  This can
be explained if the weaker objects contain no radio jets, or jets that are not
powerful enough to blast out of the vicinity of the active nucleus.  Then,
the radio emission is hidden either behind the nuclear torus or behind a
relatively compact narrow-line region, and can be free-free absorbed.  An 
alternative possibility is that low-luminosity Seyferts undergo accretion 
with an ADAF, whose radio core has an inverted spectrum.}

\end{enumerate}

We conclude that all the properties of the Palomar Seyfert sample are
consistent with the unified scheme for Seyferts {\it if} two conditions hold.
First, there is some threshold level of Seyfert activity, below which the
probability of a radio jet emerging from the nuclear region is small.
Second, the lowest luminosity radio sources either are significantly 
free-free absorbed by ionized gas or are generated by highly 
sub-Eddington accretion flows.

\acknowledgments

We thank Neal Miller for assistance with the statistical tests, Mark Claussen 
for advice on plotting the figures, and Henrique Schmitt for useful 
discussions.  We thank the referee, Ski Antonucci, for useful suggestions
that helped clarify the paper.  The National Radio Astronomy Observatory 
is a facility of the National Science Foundation, operated under cooperative
agreement by Associated Universities, Inc.  This research has made use of the NASA/IPAC 
Extragalactic Database (NED) which is operated by the Jet Propulsion Laboratory, 
California Institute of Technology, under contract with the National Aeronautics 
and Space Administration.  In addition, this research has made use of NASA's
Astrophysics Data System Abstract Service.  The work of L.~C.~H. is partly 
funded by NASA grants HST-GO-06837.04-A, HST-AR-07527.03-A, and 
HST-AR-08361.02-A, awarded by the Space Telescope Science Institute, which is 
operated by AURA, Inc., under NASA contract NAS5-26555.

\clearpage

\begin{deluxetable}{ccc}
\tablecolumns{3}
\tablewidth{140pt}
\tablecaption{Optical Luminosity Function for Palomar Seyferts}
\tablehead{
\colhead{$M_{B_T}$\tablenotemark{a}}&\colhead{$\log \phi $}&
\colhead{No.} \\
\colhead{(mag)} &\colhead{(Mpc$^{-3}$mag$^{-1}$)}& \\ }
\startdata
$-18.25$&$-3.17^{+0.23}_{-0.56}$&\ 2 \\
$-18.75$&\nodata&\ 0 \\
$-19.25$&$-3.18^{+0.14}_{-0.22}$&\ 7 \\
$-19.75$&$-3.59^{+0.16}_{-0.27}$&\ 5 \\
$-20.25$&$-3.57^{+0.13}_{-0.17}$& 11 \\
$-20.75$&$-3.92^{+0.13}_{-0.20}$&\ 9 \\
$-21.25$&$-4.43^{+0.19}_{-0.36}$&\ 4 \\
$-21.75$&$-5.08^{+0.20}_{-0.39}$&\ 3 \\
\enddata
\tablenotetext{a}{Total (nucleus + host galaxy) magnitude.  Bins
outside the tabulated range include no more than one galaxy each,
and are not shown.}
\label{tab:optlum}
\end{deluxetable}

\begin{deluxetable}{cccccc}
\tablecolumns{6}
\tablewidth{0pc}
\tablecaption{Radio Luminosity Functions for Palomar Seyferts}
\tablehead{
&\multicolumn{2}{c}{6~cm} && \multicolumn{2}{c}{20~cm} \\ 
\cline{2-3} \cline{5-6} \\ 
\colhead{$\log P $\tablenotemark{a}}&\colhead{$\log \phi $}&
\colhead{No.}&& \colhead{$\log \phi $}&\colhead{No.} \\ 
\colhead{(W~Hz$^{-1}$)}&\colhead{(Mpc$^{-3}$mag$^{-1}$)}&&&
\colhead{(Mpc$^{-3}$mag$^{-1}$)}&  \\ }
\startdata
19.0&$-4.32^{+0.20}_{-0.39}$&3.3&&\nodata&\nodata \\
19.4&$-3.38^{+0.21}_{-0.42}$&7.7&&$-3.57^{+0.28}_{-1.07}$&2 \\
19.8&$-4.09^{+0.22}_{-0.48}$&4&&$-4.54^{+0.21}_{-0.42}$&3 \\
20.2&$-3.94^{+0.19}_{-0.36}$&4&&$-3.87^{+0.19}_{-0.37}$&4 \\
20.6&$-4.81^{+0.23}_{-0.53}$&2&&$-4.26^{+0.21}_{-0.41}$&4 \\
21.0&$-4.21^{+0.19}_{-0.37}$&6&&$-4.22^{+0.20}_{-0.38}$&5 \\
21.4&$-4.31^{+0.20}_{-0.42}$&6&&$-5.38^{+0.20}_{-0.38}$&3 \\
21.8&$-4.62^{+0.23}_{-0.53}$&2&&$-4.34^{+0.22}_{-0.45}$&4 \\
22.2&\nodata&0&&$-4.49^{+0.20}_{-0.38}$&3 \\
\enddata
\tablenotetext{a}{Bins outside the tabulated range include no 
more than one galaxy each, and are not shown.}
\label{tab:radlum}
\end{deluxetable}

\clearpage

\begin{deluxetable}{llcl}
\tablecolumns{4}
\tablewidth{0pc}
\tablecaption{Summary of Statistical Tests}
\tablehead{
\colhead{Independent\hfil}&\colhead{Dependent\hfil}&\colhead{Statistical}&
\colhead{Section\hfil} \\ 
\colhead{Variable\hfil}&\colhead{Variable\hfil}&
\colhead{Significance\tablenotemark{a}}& \colhead{} \\ }
\startdata
Seyfert type&Distance&54\%&3.1 \\
Seyfert type&Core 20 cm power&0.9\%&3.4.1 \\
Seyfert type&Total 20 cm power&0.9\%&3.4.1 \\
Seyfert type&Core 6 cm power&2.0\%&3.4.1 \\
Seyfert type&Total 6 cm power&1.3\%&3.4.1 \\
Seyfert type&\oiii\ luminosity&15\%&3.5 \\
Seyfert type&Total 20 cm/\oiii\ ratio&0.9\%&3.5 \\
Seyfert type&20 cm/\oiii, strong \oiii&51\%&4.2 \\
Seyfert type&20 cm/\oiii, weak \oiii&1.3\%&4.2 \\
Seyfert type&20 cm/\oiii, weak \oiii, no E gal.&0.1\%&4.2 \\
Seyfert type&Total 6 cm/\oiii\ ratio&1.2\%&3.5 \\
Seyfert type&6 cm/\oiii, strong \oiii&26\%&4.2 \\
Seyfert type&6 cm/\oiii, weak \oiii&3.1\%&4.2 \\
Seyfert type&6 cm/\oiii, weak \oiii, no E gal.&0.6\%&4.2 \\
Seyfert type&Total spectral index&29\%&3.6 \\
Seyfert type&Peak spectral index&63\%&3.6 \\
Seyfert type&Radio size, resolved&74\%&3.7 \\
Seyfert type&Radio size, all&31\%&3.7 \\
Seyfert type&Radio morphology&9.2\%&3.8 \\
Seyfert type&Radio morphology, incl. Ambiguous&2.9\%&3.8 \\
Seyfert type&$q$ = FIR/radio ratio&1.3\%&4.2 \\
Seyfert type&H$\alpha$ equivalent width&0.8\%&4.3 \\
Host-galaxy type&Seyfert type&48\%&3.2 \\
\oiii\ luminosity&Total 20 cm power&$<0.01$\% &3.5 \\
\oiii\ luminosity&Total 6 cm power&$<0.01$\% &3.5 \\
\nii\ line width&Total 20 cm power&$<0.01$\% &3.5 \\
\nii\ line width&Total 6 cm power&$<0.01$\% &3.5 \\
Total 6 cm power&Radio morphology&8.8\% &3.8 \\
Total 6 cm power&Radio morphology, incl. Ambiguous&1.4\%&3.8 \\
\enddata
\tablenotetext{a}{The statistical significance of differences between
the two Seyfert types is generally defined as the probability that the
distributions of the dependent variable would differ by the observed
amount if the two Seyfert types were drawn from the same parent population.}
\label{tab:statsumm}
\end{deluxetable}

\clearpage 
\begin{deluxetable}{llcccc}
\tablecolumns{6}
\tablewidth{0pc}
\tablecaption{Seyfert Samples Imaged in the Radio at Arcsecond Resolution}
\tablehead{
\colhead{Name}&\colhead{Selection}&\colhead{No.}&
\colhead{Observations}&\colhead{Resolution}&\colhead{References} \\ }
\startdata
Markarian&UV excess&29&VLA: 6, 20 cm&0\farcs4--1\farcs3 &1 \\
Distance-limited&Heterogeneous&57&VLA: 6, 20 cm&0\farcs4--1\farcs3 &2 \\
CfA&Optical&47&VLA: 3.6 cm&0\farcs2--2\farcs0 &3 \\
Southern&Heterogeneous&29&VLA, ATCA: 3.5, 6 cm&1\farcs0 &4 \\
12-$\mu$m&Infrared&107&VLA: 3.6 cm&0\farcs25 &5  \\
60-$\mu$m&Infrared&75&VLA: 3.6 cm&0\farcs25 &6 \\
Palomar&Optical&45&VLA: 6, 20 cm&1\farcs1 &7 \\
\enddata
\tablerefs{
(1) Ulvestad \& Wilson 1984a;
(2) Ulvestad \& Wilson 1984b, 1989;
(3) Kukula et al. 1995;
(4) Morganti et al. 1999;
(5) Thean et al. 2000, 2001;
(6) Kinney et al. 2000, Schmitt et al. 2001a, 2001b;
(7) Ho \& Ulvestad 2001, this paper.}
\label{tab:samples}
\end{deluxetable}

\clearpage
\end{document}